\documentclass[11pt]{article}
\pdfoutput=1
\usepackage{hyperref}
\usepackage{graphicx}
\usepackage{graphics}
\usepackage{dsfont}
\usepackage{epsfig}
\usepackage{amsmath,amssymb,amsthm,amscd}

\usepackage{adjustbox}

\newtheorem{proposition}{Proposition}
\newcommand{\be}{\begin{equation}}
\newcommand{\ee}{\end{equation}}
\newcommand{\ba}{\begin{aligned}}
\newcommand{\ea}{\end{aligned}}

\newcommand{\rank}{\,\mathrm{rank}\,}

\newcommand{\sn}{\,\mathrm{sn}\,}

\newcommand{\ri}{\mathrm{i}}

\numberwithin{equation}{section}

\begin{document}
\begin{titlepage}
{}~ \hfill\vbox{ \hbox{} }\break

\begin{flushright}
    USTC-ICTS/PCFT-23-33\\
    KIAS-Q23022
\end{flushright}
\vskip 1.5 cm

\begin{center}
{\Large  \bf  
Schur indices for $\mathcal{N}=4$ super-Yang-Mills
 \\ \vskip 0.2cm 
 with more general gauge groups}
\end{center}

\vskip 0.5 cm

\renewcommand{\thefootnote}{\fnsymbol{footnote}}
\vskip 30pt \centerline{ {\large \rm 
Bao-ning Du\footnote{baoningd@mail.ustc.edu.cn}, 
Min-xin Huang\footnote{minxin@ustc.edu.cn}, Xin Wang\footnote{wxin@kias.re.kr}  
} } \vskip .5cm  \vskip 20pt

\begin{center}
{$^*$ Max Planck Institute for Mathematics,   \\ \vskip 0.1cm
Vivatsgasse 7, D-53111 Bonn, Germany}
 \\ \vskip 0.3 cm
{$^\dagger$ Interdisciplinary Center for Theoretical Study,  \\ \vskip 0.1cm  University of Science and Technology of China,  Hefei, Anhui 230026, China} 
 \\ \vskip 0.3 cm
{$^\dagger$ Peng Huanwu Center for Fundamental Theory,  \\ \vskip 0.1cm  Hefei, Anhui 230026, China} 
 \\ \vskip 0.3 cm
{$^\ddagger$ Quantum Universe Center, Korea Institute for Advanced Study  \\ \vskip 0.1cm Hoegiro 85, Seoul 02455, Korea}
\end{center}

\setcounter{footnote}{0}
\renewcommand{\thefootnote}{\arabic{footnote}}
\vskip 50pt
\begin{abstract}
We study the unflavored Schur indices in the $\mathcal{N}=4$ super-Yang-Mills theory for the $B_n,C_n,D_n, G_2$ gauge groups. We explore two methods, namely the character expansion method and the Fermi gas method, to efficiently compute the $q$-series expansion of the Schur indices to some high orders. Using the available data and the modular properties, we are able to fix the exact formulas for the general gauge groups up to some high ranks and discover some interesting new features. We also identify some empirical modular anomaly equations, but unlike the case of $A_n$ groups, they are quite complicated and not sufficiently useful to fix exact formulas for gauge groups of arbitrary rank.

\end{abstract}

\end{titlepage}
\vfill \eject


\newpage

\baselineskip=16pt

\tableofcontents

\section{Introduction}
As a type of topological invariants, Witten index is a very powerful non-perturbative tool for supersymmetric theories \cite{Witten:1982df}. In the context of superconformal quantum field theories, such indices were first constructed in \cite{Romelsberger:2005eg, Kinney:2005ej}. For theories with a Lagrangian description, the $d$-dimensional superconformal index can be computed by path integral formalism as the supersymmetric partition function on $S^1 \times S^{d-1}$. If an effective Lagrangian flows to a superconformal fixed point in the UV or IR, we can also compute the superconformal index at the fixed point from the Lagrangian theory. There are many important applications, e.g. providing quantitative tests of field theory dualities, and counting holographic dual black hole microstates. We will focus on the 4d case, where the superconformal index can be computed with an integral over the Cartan generators of the gauge algebra, counting gauge invariant operators. For reviews see e.g. \cite{Rastelli:2016tbz, Gadde:2020yah}.

For theories with extended supersymmetry, a particular specialization of the 4d superconformal index, known as the Schur index \cite{Gadde:2011uv}, has some further nice mathematical properties. For example, in some cases it can be computed from the $q$-deformed 2d Yang-Mills \cite{Gadde:2011ik}, or the vacuum character of a corresponding chiral algebra \cite{Beem:2013sza,Beem:2017ooy}. The Schur index has been studied extensively, e.g. \cite{Bourdier:2015wda, Pan:2021mrw, Beem:2021zvt}. For the case of $\mathcal{N}=4$ supersymmetry, besides a universal fugacity parameter denoted as $q$, the Schur index may have an extra flavor fugacity from the symmetry $SU(2)_F\subset SU(4)_R$. We will continue the study of unflavored Schur index in the 
$\mathcal{N}=4$ super-Yang-Mills theory by one of the authors in \cite{Huang:2022bry}, generalizing the results for $A_n$ gauge groups to more gauge groups $B_n,C_n,D_n, G_2$. A modular anomaly equation was proposed and proved in \cite{Huang:2022bry}, inspired by similar equations in topological string theories \cite{Bershadsky:1993cx, Minahan:1997ct}. We also explore the modular anomaly equations in this paper and find that they are much more complicated comparing to the case of $A$-type gauge groups. Some other recent studies relating to Schur index are \cite{Zheng:2022zkm, Eleftheriou:2022kkv, Hatsuda:2022xdv, Hatsuda:2023iwi, Guo:2023mkn, ArabiArdehali:2023bpq,Hatsuda:2023imp}. The AdS/CFT correspondence for $BCD$ types of gauge groups was described in \cite{Witten:1998xy}.

The paper is organized as follows.  In Section \ref{chapter2} we introduce the general formulas for the Schur index and discuss the modular properties for various gauge groups. We discuss how to compute the $q$-series expansion and use the modular property to fix the exact formulas for the case of $G_2$ gauge group.  In Section \ref{SecCharacter} we explore the character expansion method to compute the $q$-series expansion for the $BCD$-type gauge groups.  In Section \ref{SecFermi} we explore the Fermi gas method. We find that the method is more efficient for $D$-type gauge groups than $BC$-type gauge groups.  In Section \ref{modular anomaly equations} we use the perturbative results as well as some empirical patterns to fix the exact modular formulas up to some high ranks, and discuss the modular anomaly equations. Finally, in Section \ref{SecDiscuss} we discuss the main results and potential future research. In Appendix \ref{app:A} we provide our convention for elliptic functions and modular forms. In Appendix \ref{Seccongruence} we discuss the generators of modular forms of some relevant congruence subgroups. In Appendix \ref{results} we list the main results of the exact modular formulas.

\section{The Schur indices: some general properties} \label{chapter2}
The 4d $\mathcal{N}=2$ superconformal index is a type of Witten index \cite{Kinney:2005ej}, as a trace over the Hilbert space on $S^3$ in the radial quantization. For a theory with a weakly coupled Lagrangian description, the index is computed explicitly as a matrix integral \cite{Gadde:2011ik}. In this paper, we consider the Schur limit of the superconformal index, known as the Schur index, for the 4d $\mathcal{N}=4$ super-Yang-Mills theories. Alternatively, the Schur index can be derived by the supersymmetric localization of $\mathcal{N}=2$ SCFT theory on $S^3\times S^1$. For the $\mathcal{N}=4$ case, it can be regarded as a $\mathcal{N}=2^{*}$ theory with massless adjoint hypermultiplet, and we have \footnote{The expression of the Schur index appears in different forms in the literature up to a factor $q^{c(G)}$. Here we use a convention such that the result has nice modular properties. See Appendix \ref{app:A} for notations of Jacobi theta functions $\theta_i(z)$ and other elliptic functions.}
\begin{equation}\label{eq:SI_def}
\begin{split}
	 \mathcal{I}_G(q)&=\frac{(-\ri)^{\rank G-\dim G}\,\eta(\tau)^{3\rank{G}}}{|W|\,\theta_4^{\rank{G}}}\oint \prod_{j=1}^{\rank{G}}da_j\prod_{\rho \in \mathcal{R}_{G}^{*}}\frac{\theta_1(\rho(a_j))}{\theta_4(\rho(a_j))}\\
  &=\frac{1}{|W|}\left(\frac{\eta(\tau)^{3}}{\theta_4}\right)^{\rank{G}}\oint \prod_{j=1}^{\rank{G}}da_j\prod_{\rho \in \mathcal{R}_{G}^{+}}\frac{\theta_1(\rho(a_j))^2}{\theta_4(\rho(a_j))^2},
\end{split}  
\end{equation}
where $|W|$ stands for the dimension of the Weyl group of the gauge group $G$, $\dim G$ is the dimension of the adjoint representation, $\rank G$ is the rank of $G$, $\mathcal{R}_{G}^{*}$ is the set of non-zero roots, and $\mathcal{R}_{G}^{+}$ is the set of positive roots. The integrals $\oint$ are performed for $a_i$ from $0$ to $2\pi \ri$, effectively eliminating the terms $e^{ma}$ with $m\in \mathbb{Z} -\{0\}$, leaving only the constant term, similar as a residue calculation.

The Schur index can be calculated by explicitly performing the integral \eqref{eq:SI_def}. It is usually difficult to directly compute the integrals of theta functions, see e.g. some examples of such elaborate efforts in \cite{Bourdier:2015wda,  Pan:2021mrw}. Instead, we will expand the integrand in terms of $q=e^{\pi \ri \tau} $ to some high order, and integrate over the gauge fugacities $a_i$, then we can use the ring of quasi-modular forms in the congruence subgroups listed in Table \ref{tab:modularform}, to fix the exact modular expressions of the Schur index. More precisely, the Schur indices can be expressed as
\begin{align} \label{tildeI}
    \mathcal{I}_G(q)=s_0^{b(G)}\cdot \widetilde{\mathcal{I}}_G(q),
\end{align}
where 
\begin{align}\label{s0}
    s_0=\frac{\theta_4(\tau)}{\eta(\tau)^3}=q^{-\frac{1}{4}}(1-2q+3q^2-6q^3+\cdots),
\end{align}
and $\widetilde{\mathcal{I}}_G(q)$ can be written in terms with the second Eisenstein series and the modular forms of the congruence subgroup $\Gamma\subset \mathrm{SL}_2(\mathbb{Z})$. The subgroups $\Gamma$ and the values of $b(G)$ are summarized in Table \ref{tab:modularform}. Some properties and generators of the relevant congruence subgroups are explained in Appendix \ref{Seccongruence}. 

\begin{table}[!h]
    \centering
    \begin{tabular}{|c|c|c|c|c|c|c|}\hline
      Group &$A_{2N-1}$&$A_{2N}$&$B_{N},C_{N}$&$D_{2N-1}$&$D_{2N}$& $G_2$ \\ \hline
       Modular group $\Gamma$&$\Gamma^0(2)$&$\mathrm{SL}_2(\mathbb{Z})$&\multicolumn{3}{|c|}{$\Gamma(2)$}& $\Gamma_0(6)\cap\Gamma^0(2)$ \\
        \hline
        $b(G)$ & 0&1 & $N$& $2N-3$&$2N$ & 2 \\
        \hline
    \end{tabular}
    \caption{The modular groups for $\mathcal{N}=4$ Schur indices. The modular group of $G_2$ can be also written as $\Gamma_0(3)\cap \Gamma(2)$.}
    \label{tab:modularform}
\end{table}

For example, for the $U(N)$ group, we have
\begin{align}
    \mathcal{I}_{U(N)}(q)=\frac{\eta(\tau)^{3N}}{N!\,\theta_4^{N}}\oint\prod_{j=1}^{N}da_j\prod_{1 \leq i<j\leq N}\frac{\theta_1(a_i-a_j)^2}{\theta_4(a_i-a_j)^2},
\end{align}
from which we can solve
\begin{equation}
    \mathcal{I}_{U(1)}=s_0^{-1}=\frac{\eta^3}{\theta_4}.
\end{equation}
The $SU(N)$ Schur index can be solved from $U(N)$ Schur index, by factoring out the $U(1)$ contribution, we have
\begin{align}
    \mathcal{I}_{SU(N)}(q)=\mathcal{I}_{U(N)}(q)/\mathcal{I}_{U(1)}(q)=\frac{\eta(\tau)^{3(N-1)}}{N!\,\theta_4^{N-1}}\oint\prod_{j=1}^{N}da_j\prod_{1 \leq i<j\leq N}\frac{\theta_1(a_i-a_j)^2}{\theta_4(a_i-a_j)^2}.
\end{align}
One of the integrals in $\oint\prod_{j=1}^{N}da_i$ is actually trivial so we have indeed $N-1$ integrals corresponding to the rank. We also use $\mathcal{I}_{A_{N-1}}$ to denote the Schur index for the $SU(N)$ gauge group. For example for the low ranks, we have
\begin{align}
    \mathcal{I}_{A_1}&=\frac{1}{2}E_2+\frac{1}{24}(\theta_2^4+\theta_3^4),\\
   \widetilde{ \mathcal{I}}_{A_2}&=\frac{1}{2}E_2+\frac{1}{24}.
\end{align}
The formulas for $A_N$ of arbitrary rank have been determined \cite{Beem:2021zvt, Pan:2021mrw, Huang:2022bry}.

Before focusing on the main examples of $BCD$-type gauge groups, we consider an isolated but still non-trivial example of the $G_2$ gauge group.  Based on the symmetry of the Dynkin diagram, we propose the congruence subgroup of Schur index to be $\Gamma_0(6)\cap\Gamma^0(2)$, as shown in Table \ref{tab:modularform}. In this case, the rank is quite small, and there are only two integrals to perform. We can compute up to a very high order in the $q$-series expansion and fix the exact formula in terms of the generators of the congruence subgroup with some redundant checks. The formula is
\begin{align} \label{formulaG2}
    \mathcal{I}_{G_2}(q)&=\frac{\eta(\tau)^{6}}{12\,\theta_4^{2}}\oint\prod_{j=1}^{2}da_j\cdot\frac{\theta_1(a_1)^2\theta_1(a_2)^2\theta_1(3a_1-a_2)^2}{\theta_4(a_1)^2\theta_4(a_2)^2\theta_4(3a_1-a_2)^2}\\
    &=  \left(\frac{\theta_4 }{\eta^3} \right)^{2}\left(\mathcal{I}_{A_1}^2-\frac{1}{288}(8E_2^{(3)}+\theta_2^4-8\theta_3^4+3\theta_2^4\theta_3^4)\right),
\end{align}
where $E_2^{(3)}$ is a weight two modular form defined in \eqref{eq:E3}.

As expected from the famous S-duality of $\mathcal{N}=4$ super-Yang-Mils theory, we can check that the Schur indices for the $B_N$ and $C_N$ groups are the same up to some low orders in the $q$-series expansion. The tests can be pushed to much higher orders with the more efficient methods in the next two sections. It is well known that the $B_N$ and $C_N$ root systems have the same Weyl group and similar structures \cite{koike1987young}. For low ranks we may find an explicit isomorphic transformation between the roots of these two groups. For example for the simple case of $N=2$, the Dynkin diagrams of $B_2$ and $C_2$ are actually the same. In our parametrization explained in more detail later (\ref{BCDroots}), their roots are expressed in terms of simple roots as follows
\be\ba
	B_2&:~~\mathcal{R}_{B_2}^*=\{ \pm a_1,\pm a_2,\pm(a_1\pm a_2)  \},\\
	C_2&:~~\mathcal{R}_{C_2}^*=\{ \pm 2a'_1,\pm 2a'_2,\pm(a'_1\pm a'_2)  \}.
\ea\ee
We can set a relation between simple roots $a_1=a'_1+a'_2,a_2=a'_1-a'_2$, transforming the roots of $B_2$ to $C_2$. Since the computations of the Schur index pick out the constant terms of the simple roots, they are the same $\mathcal{I}_{B_2}=\mathcal{I}_{C_2}$ in this case. However, for general rank $N$, we are not aware of a universal mathematical proof of the non-trivial identity $\mathcal{{I}}_{B_N} = \mathcal{{I}}_{C_N}$.

\section{Character expansion method} \label{SecCharacter}

From the definition of the Witten index, the Schur index can be written as an integral of plethystic exponential function \cite{Gadde:2011uv}, which in our context can be obtained by rewritings of the Jacobi theta functions in (\ref{eq:SI_def}). They can then be expanded as characters of the gauge groups, see e.g. some early works \cite{Dolan:2007rq, Dutta:2007ws}. The formalism is nicely applied to the $BCD$ types of gauge groups in a recent paper \cite{Sei:2023fjk}, which provides explicit formulas in terms of sums over 2D Young diagrams of integer partitions. 

The Schur index for the (mass deformed)  4d $\mathcal{N}=4$ super-Yang-Mills theory with gauge group $G$ has the expression
\begin{equation}\label{characterindex}
    \mathcal{I}_G(q;m)=q^{c(G)} \int [da]\mathrm{PE}\left[\chi^G_{\mathrm{\mathbf{Adj}}}(e^{\alpha})\left(i_V(q)+i_H(q)(e^m+e^{-m})\right)\right], 
\end{equation}
where PE denotes the plethystic exponential function, $i_V(q)$ and $i_H(q)$ are the single letter indices for the vector multiplet and hypermultiplet that are defined as
\begin{equation}
    i_V(q)=\frac{-2q^2}{1-q^2},\quad\quad i_H(q)=\frac{q}{1-q^2}.
\end{equation}
Here $\chi^G_{\mathrm{\mathbf{Adj}}}(e^{\alpha})$ denotes the character of the gauge group $G$ in the adjoint representation, it is defined as the sum over the roots of $G$
\begin{align}
    \chi^G_{\mathrm{\mathbf{Adj}}}(e^{\alpha})=\sum_{\rho\in\mathcal{R}_{G}}e^{\rho(a_i)}.
\end{align}
$[da]$ is the Haar measure defined as
\begin{align}
    [da]=\frac{1}{|W|}\prod_{i}\frac{da_i}{2\pi \ri}\cdot \prod_{\rho \in\mathcal{R}_{G}^+}\left(1-e^{\rho(a)}\right)^2,
\end{align}
where we used $\mathcal{R}_{G}$ and $\mathcal{R}_{G}^+$ denote the roots and positive roots of the lie algebra corresponds to $G$.

For the massless case, the total single letter index is $i_{\mathcal{N}=4} (q)= \frac{2q}{1+q}$. The integral in (\ref{characterindex}) can be expanded a series of $q$,  starting from a constant term as $1+\mathcal{O}(q)$. The factor $q^{c(G)}$ can be determined by comparing with the formula (\ref{eq:SI_def}), counting such powers of $q$ from the $\eta, \theta_1$ functions.

The formula (\ref{characterindex}) can be computed using the properties of character for the $BCD$ types of groups \cite{Sei:2023fjk, Hatsuda:2023iwi}. They can be expressed by the following formulas respectively
\begin{align}
 \label{character_Bn} \mathcal{I}_{B_N}&= q^{c(G)} \sum_{\lambda}\Big[ \frac{f_\lambda(q)}{z_\lambda}\frac{1}{2^{l(\lambda)}} \sum_{\tilde{\lambda}\in \text{Ev}(\lambda) }(-1)^{l(\tilde{\lambda})} \sum_{\mu\in R^r_{2N+1}(2|\lambda|)}\chi^S_\mu(\tilde{\lambda}) \Big], \\ 
  \label{character_Cn} \mathcal{I}_{C_N}&=q^{c(G)} \sum_{\lambda}\Big[ \frac{f_\lambda(q)}{z_\lambda}\frac{1}{2^{l(\lambda)}} \sum_{\tilde{\lambda}\in \text{Ev}(\lambda) } \sum_{\mu\in R^c_{2N}(2|\lambda|)}\chi^S_\mu(\tilde{\lambda}) \Big],\\
 \label{character_Dn}	\mathcal{I}_{D_N}&= q^{c(G)} \sum_{\lambda}\Big[ \frac{f_\lambda(q)}{z_\lambda}\frac{1}{2^{l(\lambda)}} \sum_{\tilde{\lambda}\in \text{Ev}(\lambda) } (-1)^{l(\tilde{\lambda})}  \sum_{\mu\in R^r_{2N}(2|\lambda|)\cup W^r_{2N}(2|\lambda|)}\chi^S_\mu(\tilde{\lambda}) \Big],  \\
f_\lambda(q)&:=\prod_{i=1}^{l(\lambda)}i_{\mathcal{N}=4}(q^{\lambda_i}),~~~~z_{\lambda}:=\prod_{i=1}^\infty k_i!i^{k_i}.
\end{align}
We use the notations in \cite{Sei:2023fjk}, explained in the followings. The sums are over integer partitions $\lambda$, depicted by 2D Young tableaux as non-increasing sequences $\lambda_1\geq \lambda_2 \geq \cdots $. The usual notations $l(\lambda), |\lambda|$ denote the number of rows and boxes and $\lambda^T$ denote the transpose of the Young tableau. $k_m (\lambda)=\lambda^T_m- \lambda^T_{m+1}$ is the number of $m$'s in the partition. $\text{Ev}(\lambda), R^r_n(|\lambda|),W^r_{n}(|\lambda|)$ and $R^c_{n}(|\lambda|)$ are the set of partitions defined by:
\begin{itemize}
    \item $\text{Ev}(\lambda)$: a set of partitions that is obtained by replacing $\lambda_i$ with $2\lambda_i$ or $\lambda_i,\lambda_i$ for all  $i$'s,
    \item $R^r_{n}(|\lambda|):=\{ \mu |~  l(\mu)\le n , |\mu|=|\lambda|,\forall i(\mu_i ~\text{is even})  \}$,
    \item $W^r_{n}(|\lambda|):=\{ \mu|~   l(\mu) = n,  |\mu|= |\lambda|,  \forall i(\mu_i ~\text{is odd})  \}$,
    \item  $R^c_{n}(|\lambda|):=\{ \mu |~  l(\mu)\le n , |\mu|=|\lambda|,\forall i(\mu^T_i ~\text{is even})  \}$ .
\end{itemize} 
Finally, $\chi_\mu^S(\lambda)$ is the character of the symmetric group $S_{|\lambda|}$, defined for $|\mu|=|\lambda|$. It can be computed by the Frobenius method, as the coefficient of $\prod_{i=1}^{l(\mu)}  x_i ^{\mu_i +l(\mu)-i}$ in the following expression 
\be
\prod_{1\leq i<j\leq l(\mu)} (x_i-x_j)  ~\cdot ~ \prod_{j=1}^{\lambda_1} (\sum_{i=1}^{l(\mu)} x_i^j)^{k_j(\lambda)}.
\ee

The character expansion method is much more efficient than directly computing (\ref{eq:SI_def}). For example, we can check the S-duality relation $\mathcal{{I}}_{B_N} = \mathcal{{I}}_{C_N}$ to much higher orders in $q$-series expansion.

When $N$ is large, we can compute to further high orders by comparing with the $N\sim \infty$ result, similar as in the more familiar $A_N$ case. It was pointed out in \cite{Sei:2023fjk} that for the adjoint representation, the large $N$ results are the same for $BCD$ types of gauge groups (omitting the rather trivial factor $q^{c(G)}$ in our convention), and a simple expression was proposed there 
\be
\mathcal{I}_{\infty}=  \sum_{\lambda}\Big[ \frac{f_\lambda(q)}{z_\lambda}\frac{1}{2^{l(\lambda)}} \sum_{\tilde{\lambda}\in \text{Ev}(\lambda) } (-1)^{l(\tilde{\lambda})}  \prod_{m=1}^{\infty} a_{m, k_m(\tilde{\lambda})}  \Big],
\ee
where the coefficients are determined by the  initial  values and recursion relations 
\be \ba
& a_{m,n} = m(n-1) a_{m,n-2},  ~~~~~~~~~~~~ a_{m,0}=1, a_{m,1}=0,  ~~~ \text{if} ~m~ \text{is odd}; \\
 &a_{m,n} = a_{m,n-1}+m(n-1)a_{m,n-2},  ~~ a_{m,0}=a_{m,1}=1, ~~~ \text{if} ~m~ \text{is even}.
\ea \ee
The calculation of $\mathcal{I}_{\infty}$ is much more efficient than the finite $N$ formulas (\ref{character_Bn}, \ref{character_Cn}, \ref{character_Dn}), and one can easily compute to much higher order. It is easy to see when $n\geq |\lambda|$, the sets are the same $ R^r_{n}(2|\lambda|) = R^r_{\infty}(2|\lambda|)$, so $\mathcal{{I}}_{B_N}$ agrees with $\mathcal{I}_{\infty}$ up to the order $q^{2N+1}$, i.e. 
\be
\mathcal{{I}}_{B_N} = \mathcal{{I}}_{C_N} = q^{c(G)} [\mathcal{I}_{\infty} +\mathcal{O}(q^{2N+2})]. 
\ee 
On the other hand, since $W^r_{n}(|\lambda|)= W^r_{\infty}(|\lambda|)= \O$ for $n\geq |\lambda|+1$ but $W^r_{|\lambda|}(|\lambda|)\neq \O$, the available data from $\mathcal{I}_{\infty}$ for $D$-type group is much fewer, only up to the order $q^{N-1}$, i.e. 
\be
\mathcal{{I}}_{D_N}  = q^{c(G)} [\mathcal{I}_{\infty} +\mathcal{O}(q^{N})]. 
\ee 
Some examples are the followings. Omitting the factor $q^{c(G)}$, we have 
\be\ba
 \mathcal{I}_{\infty}(q)&= 1 +3q^2 - 4q^3 + 15q^4 - 24q^5 + 62 q^6 - 120q^7 + 270 q^8 + \mathcal{O}(q^{9}), \\
    \mathcal{I}_{B_4}(q)= \mathcal{I}_{C_4}(q)&\sim \underline{1 +3q^2 - 4q^3 + 15q^4 - 24q^5 + 62 q^6 - 120q^7 + 270 q^8} + \mathcal{O}(q^{9}),\\
    \mathcal{I}_{B_3}(q)= \mathcal{I}_{C_3}(q)&\sim \underline{1 +3q^2 - 4q^3 + 15q^4 - 24q^5 + 62 q^6 - 120q^7} + 255 q^8 + \mathcal{O}(q^{9}),\\
    \mathcal{I}_{D_4}(q)&\sim \underline{1 +3q^2 - 4q^3} + 20q^4 - 32q^5 + 86 q^6 - 176q^7 + 415 q^8 + \mathcal{O}(q^{9}),\\ 
    \mathcal{I}_{D_3}(q)&\sim\underline{1 +3q^2}  + 9q^4 - 6q^5 + 22 q^6 - 18q^7 + 51 q^8 + \mathcal{O}(q^{9}),
\ea\ee
where the underlines denote the agreements with $ \mathcal{I}_{\infty}(q)$.

\section{Fermi gas method} \label{SecFermi}
The Fermi gas approach provides a powerful method to calculate the matrix integrals in various contexts \cite{Marino:2011eh,Klemm:2012ii,Hatsuda:2012dt,Grassi:2014zfa,Bonelli:2016idi}. It was first used in \cite{Bourdier:2015wda} to calculate the $\mathcal{N}=4$ Schur indices of unflavored $SU(N)$ cases and was later generalized to the flavored cases \cite{Hatsuda:2022xdv} and cases with line defects \cite{Drukker:2015spa,Horikoshi:2016hds,Hatsuda:2023iwi}. The purpose of this section is to generalize the Fermi gas approach to the $\mathcal{N}=4$ Schur indices with the $SO(N)$ and $Sp(N)$ gauge groups.

The key ingredient of the Fermi gas approach is the elliptic generalization of the \emph{Cauchy determinant formula}, that the integrand of the integral can be expressed as a determinant of a density matrix $\rho(a_i,a_j)$, for the $U(N)$ theory, we have \cite{Bourdier:2015wda}
\begin{equation}
\begin{split}\label{eq:cauchyA}
    \prod_{1\leq i<j\leq N}\frac{\theta_1(a_i-a_j)^2}{\theta_4(a_i-a_j)^2}&=f_N(\tau)\det_{1\leq i,j\leq N}\rho(a_i,a_j),
\end{split}
\end{equation}
where
\begin{equation}
    \rho(a_i,a_j)=\frac{\theta_2(a_i-a_j)}{\theta_4(a_i- a_j)}.
\end{equation}
The coefficient $f_N(\tau)$ is a simple function defined by
\begin{equation}
    f_N(\tau)=\begin{cases}
        \frac{\theta_4^{N}}{\theta_3^{N}},\quad & \text{if $N$ is even},\\
        \frac{\theta_4^{N}}{\theta_2\theta_3^{N-1}},\quad & \text{if $N$ is odd}.
    \end{cases}
\end{equation}
Equation \eqref{eq:cauchyA} is a special case of the Frobenius's determinant formula \cite{MR1579913,MR2178686,MR2249536}, the $\theta_2(a_i-a_j)$ in the numerator of $\rho(a_i,a_j)$ can be replaced by either $\theta_1(a_i-a_j)$ or $\theta_3(a_i-a_j)$ with a different prefactor $f_N(\tau)$, due to different specializations of the Frobenius's determinant formula.

By using the determinant formula \eqref{eq:cauchyA} and the Leibniz formula for the determinant, the index can be written as
\begin{equation}\label{eq:ZN}
  \mathcal{I}_{U(N)}(q)\sim Z(N),\qquad  Z(N)=\frac{1}{N!}\sum_{\sigma\in S_N}(-1)^{\epsilon(\sigma)}\int \prod_i \frac{\theta_2\left(a_i-a_{\sigma(j)}\right)}{\theta_4\left(a_i- a_{\sigma(j)}\right)} da_i,
\end{equation}
where $S_N$ is the permutation group of $N$ elements, and $\epsilon(\sigma)$ is the signature of the permutation $\sigma$. Equation \eqref{eq:ZN} can be treated as the canonical partition function of an ideal Fermi gas, it can be written as a sum over conjugacy classes of the permutation group. Define the spectral trace
\begin{align}\label{eq:Zell}
    Z_{\ell}=\int_{0}^{2\pi \ri} d a_1\cdots da_{\ell} \rho(a_1,a_2)\rho(a_2,a_3)\cdots\rho(a_{\ell-1},a_{\ell})\rho(a_{\ell},a_1),
\end{align}
then the partition function is given by
\begin{equation}\label{eq:ZNZl}
    Z(N)=\sum_{m_{\ell}}^{\prime}\prod_{\ell}\frac{Z_{\ell}^{m_{\ell}}(-1)^{m_{\ell}+1}}{m_{\ell}!l^{m_{\ell}}},
\end{equation}
where the prime denotes a sum over the conjugacy class, specified by a set of non-negative integers $m_{\ell}$ that satisfy $\sum_{\ell=1}^{N}\ell m_{\ell}=N.$ The great benefit of the Fermi gas approach is that the integral representation of $Z_{\ell}$ is relatively easy to compute, it has a close form expression 
\begin{equation}
    Z_{\ell}=\sum_{n\in\mathbb{Z}}\left(\frac{1}{q^{n-\frac{1}{2}}+q^{-(n-\frac{1}{2})}}\right)^{\ell}.
\end{equation}
So, the close-form expressions of the partition function $Z(N)$, and then the Schur index $\mathcal{I}_{U(N)}(q)$ can be written down according to \eqref{eq:ZNZl} and \eqref{eq:ZN} respectively.

\subsection{Schur indices for $BCD$-type gauge groups}
In this subsection, we generalize the Fermi gas approach to $BCD$-type gauge groups. The nontrivial roots for $B_N$, $C_N$ and $D_N$ groups are
\begin{align} \label{BCDroots}
    \mathcal{R}_{B_N}^{*}&=\{\pm a_j|1\leq j\leq N\}\cup\{\pm a_i \pm a_j|1\leq i<j\leq N\},\nonumber \\
    \mathcal{R}_{C_N}^{*}&=\{\pm 2a_j|1\leq j\leq N\}\cup\{\pm a_i \pm a_j|1\leq i<j\leq N\},\\
    \mathcal{R}_{D_N}^{*}&=\{\pm a_i \pm a_j|1\leq i<j\leq N\}, \nonumber 
\end{align}
respectively. Their roots are almost the same, but for $B$- and $C$-type groups, there are additional ``diagonal roots" $\{\pm \delta a_j|1\leq j\leq N\}$ where $\delta=1$ when the gauge group is $B_N$ and $\delta=2$ when the gauge group is $C_N$. In the spirit of \cite{MR2178686,MR2249536}, we propose two propositions that will be shown to be useful for later calculations:
\begin{proposition}\label{Proposition1}
For the $D_{2N}$ group, we have the determinant formula:
    \begin{align}\label{eq:cauchyDv2}
    \prod_{1\leq i<j\leq 2N}\frac{\theta_1(a_j\pm a_i)^2}{\theta_4(a_j\pm a_i)^2}=\det_{1\leq i,j\leq 2N}\left(\frac{\theta_1(a_j\pm a_i)}{\theta_4(a_j \pm a_i)}\right).
\end{align}
For the $D_{2N-1}$ group, we have the determinant formula:
\begin{align}\label{eq:cauchyDv2_odd}
   \prod_{1\leq i<j\leq 2N-1}\frac{\theta_1(a_j\pm a_i)^2}{\theta_4(a_j\pm a_i)^2}=\det_{1\leq i,j\leq 2N}\left(\rho_{ij}\right),
\end{align}
where $\rho_{ij}$ is an anti-symmetric matrix, with $\rho_{ii}=0,\,\rho_{ij}=-\rho_{ji}$ and when $i<j$,
\begin{equation}
   \rho_{ij}=\begin{cases}
       \frac{\theta_1(a_j\pm a_i)}{\theta_4(a_j \pm a_i)},\quad &\text{if } i<j\leq 2N-1,\\
       -1,\quad &\text{if } i= 2N,j<2N.
   \end{cases} 
\end{equation}
\end{proposition}\label{Proposition2}
The proof of Proposition \eqref{Proposition1} can be found in Appendix \ref{app:proof}.

As a direct consequence of Proposition \ref{Proposition1}, the determinant formulas for $BC$-type groups can be written from the determinant formulas for $D$-type groups, we have:
\begin{proposition}
For the $B_{2N}$ or $C_{2N}$ group, we have the determinant formula
    \begin{align}
    \prod_{1\leq i<j\leq 2N}\frac{\theta_1(\delta a_j)^2}{\theta_4(\delta a_j)^2}\cdot\frac{\theta_1(a_j\pm a_i)^2}{\theta_4(a_j\pm a_i)^2}=\det_{1\leq i, j\leq 2N}\left(\frac{\theta_1(\delta a_j)^2}{\theta_4(\delta a_j)^2}\cdot\frac{\theta_1(a_j\pm a_i)}{\theta_4(a_j \pm a_i)}\right).
\end{align}
where $\delta=1$ when the gauge group is $B$-type and $\delta=2$ when the gauge group is $C$-type.
For the $B_{2N-1}$ or $C_{2N-1}$ group, we have the determinant formula
\begin{align}
   \prod_{1\leq i<j\leq 2N-1}\frac{\theta_1(\delta a_j)^2}{\theta_4(\delta a_j)^2}\cdot\frac{\theta_1(a_j\pm a_i)^2}{\theta_4(a_j\pm a_i)^2}=\det_{1\leq i,j \leq 2N}\left(\rho_{ij}\right),
\end{align}
where $\rho_{ij}$ is an anti-symmetric matrix, with $\rho_{ii}=0$ and when $i\neq j$,
\begin{equation}
   \rho_{ij}=\begin{cases}
       \frac{\theta_1(\delta a_j)^2}{\theta_4(\delta a_j)^2}\cdot\frac{\theta_1(a_j\pm a_i)}{\theta_4(a_j \pm a_i)},\quad &\text{if } i<j\leq 2N-1,\\
       -\rho_{ji},&\text{if } j<i\leq 2N-1,\\
       -1,\quad &\text{if } i< 2N,j=2N,\\
       \frac{\theta_1(\delta a_j)^2}{\theta_4(\delta a_j)^2},\quad &\text{if } i= 2N,j<2N.
   \end{cases} 
\end{equation}
\end{proposition}
In the following subsections, we use these determinant formulas to perform the calculations on the integral representations of Schur indices.
\subsubsection{Even ranks}\label{sec:Evenranks}
\paragraph{\underline{$D_{2N}$}}
We want to use the Fermi gas approach to compute the Schur index for $D_{2N}$ theory, the key part is to get the result for $Z_{\ell}$ defined in \eqref{eq:Zell}. 
By using the identities of Jacobi elliptic functions:
\begin{align}
\frac{ \theta_1 ( z ) } { \theta_4 ( z ) }
= \frac{\theta_{2} }{\theta_3 } \sn (z \theta_{3}^2)=\frac{2i}{\theta_2\theta_3}\sum_{n\in\mathbb{Z}}\frac{q^{n+\frac{1}{2}}e^{(n+\frac{1}{2})z}}{1-q^{2n+1}}=\frac{2i}{\theta_2\theta_3}\sum_{n=0}^{\infty}\frac{q^{n+\frac{1}{2}}}{1-q^{2n+1}}\left(e^{(n+\frac{1}{2})z}-e^{-(n+\frac{1}{2})z}\right)\, ,
\end{align}
the density matrix for $D_{2N}$ gauge group of the Fermi gas is
\begin{align}
   \rho(a,a^{\prime}) \equiv \frac{ \theta_1 ( a+a^{\prime} )\theta_1 ( a-a^{\prime} ) } { \theta_4 ( a+a^{\prime} )\theta_4 ( a-a^{\prime} ) }=-\frac{4}{(\theta_2\theta_3)^2}\sum_{n,n^{\prime}\in\mathbb{Z}}\frac{q^{n+n^{\prime}+1}e^{(n+n^{\prime}+1)a+(n-n^{\prime})a^{\prime}}}{(1-q^{2n+1})(1-q^{2n^{\prime}+1})}.
\end{align}
By changing the parameters, we can rewrite the summation in a form that is convenient for later calculations
\begin{align}
   \rho(a,a^{\prime}) =-\frac{4}{(\theta_2\theta_3)^2}\left(\sum_{n,n^{\prime}\in2\mathbb{Z}}\frac{q^{n+1}e^{(n+1)a+n^{\prime}a^{\prime}}}{(1-q^{n+n^{\prime}+1})(1-q^{n-n^{\prime}+1})}+\sum_{n,n^{\prime}\in2\mathbb{Z}+1}\frac{q^{n+1}e^{(n+1)a+n^{\prime}a^{\prime}}}{(1-q^{n+n^{\prime}+1})(1-q^{n-n^{\prime}+1})}\right).
\end{align}
Define
\begin{align}\label{eq:integralEven}
    Z_{\ell}=\int_0^{2\pi \ri}da_1\cdots d a_l \rho(a_1,a_2)\cdots \rho(a_{\ell},a_1),
\end{align}
only the zero order part of $a_j$ contributes to the integral. One may then find a surprisingly easy solution
\begin{align}
    Z_{2\ell}&=\frac{2^{4\ell+1}}{(\theta_2\theta_3)^{4\ell}}\left(\sum_{n_j\in\mathbb{Z}}\frac{q^{2(n_1+\cdots n_{2\ell})+\ell}}{\prod_{j=1}^{\ell}(1-q^{2n_{2j-1}\pm 2n_{2j}+1})(1-q^{2n_{2j}\pm (2n_{2j+1}+1)})}\right)\nonumber\\
    &=2^{\ell+1}\left(\frac{\eta^3}{\theta_4}\right)^{-4\ell}\left(\sum_{n_j\in\mathbb{Z}_{\geq 0}}\frac{q^{2(n_1+\cdots n_{2\ell})+\ell}\prod_{j=1}^{\ell}2^{1-\delta_{0,n_{2j}}}}{\prod_{j=1}^{\ell}(1-q^{2n_{2j-1}\pm 2n_{2j}+1})(1-q^{2n_{2j}\pm (2n_{2j+1}+1)})}\right),
\end{align}
with the notation $n_{2\ell+1}=n_1$, and
\begin{align}
    Z_{2\ell+1}=0.
\end{align}
One can then compute the Schur index of $D_{2N}$ gauge group 
\begin{align}\label{eq:Sc_D2N}
    \mathcal{I}_{D_{2N}}=\frac{1}{2^{2N-1}}\left(\frac{\eta^3}{\theta_4}\right)^{2N}\sum_{m_{\ell}}^{\prime}\prod_{\ell}\frac{Z_{2\ell}^{m_{\ell}}(-1)^{m_{\ell}}}{m_{\ell}!(2l)^{m_{\ell}}},
\end{align}
where the prime denotes a sum over non-negative integers $m_{\ell}$ that satisfy $\sum_{\ell=1}^{N}\ell m_{\ell}=N.$\footnote{In the $A$-type case, the grand canonical ensemble 
\begin{align*}
    \Xi(\kappa)=1+\sum_{N=1}^{\infty}Z(N)\kappa^N=\exp\left(-\sum_{\ell=1}^{\infty}\frac{(-\kappa)^{\ell}}{\ell}Z_{\ell}\right)
\end{align*}
has a closed form expression and was used in \cite{Bourdier:2015wda} to further simplify the result. For the $D_{2N}$-type and all other types we derive in the paper, we couldn't find a closed form expression for it, so we don't use the grand canonical ensemble calculations in the paper.
}

\paragraph{\underline{$B_{2N}$ and $C_{2N}$}}
Recall that the non-zero roots for $B_{N}$ algebra are
\begin{align}
    \mathcal{R}_{B_N}^{*}=\{\pm a_j|1\leq j\leq N\}\cup\{\pm a_i \pm a_j|1\leq i<j\leq N\},
\end{align}
and for $C_{N}$ algebra are
\begin{align}
    \mathcal{R}_{C_N}^{*}=\{\pm 2a_j|1\leq j\leq N\}\cup\{\pm a_i \pm a_j|1\leq i<j\leq N\},
\end{align}
from which we have the density function
\begin{align}\label{eq:rh0BC}
   &\rho(a,a^{\prime}) \equiv \frac{ \theta_1 (\delta a )^2\theta_1 ( a+a^{\prime} )\theta_1 ( a-a^{\prime} ) } {\theta_4 (\delta a )^2 \theta_4 ( a+a^{\prime} )\theta_4 ( a-a^{\prime} ) }\nonumber\\
   &=\frac{16}{(\theta_2\theta_3)^4}\sum_{m\in\mathbb{Z}}\left(\sum_{n,n^{\prime}\in2\mathbb{Z}}\frac{c(m)q^{n+1}e^{(n+1+\delta m)a+n^{\prime}a^{\prime}}}{(1-q^{n+n^{\prime}+1})(1-q^{n-n^{\prime}+1})}+\sum_{n,n^{\prime}\in2\mathbb{Z}+1}\frac{c(m)q^{n+1}e^{(n+1+\delta m)a+n^{\prime}a^{\prime}}}{(1-q^{n+n^{\prime}+1})(1-q^{n-n^{\prime}+1})}\right),
\end{align}
where $\delta=1$ for $B$-type and $\delta=2$ for $C$-type and
\begin{align}
    c(m)=
    \begin{cases}
        \frac{m q^m}{1-q^{2m}},& \text{if } m\neq 0,\\
    -2\mathcal{I}_{A_1},              & \text{if } m=0.
    \end{cases}
\end{align}
The last line in equation \eqref{eq:rh0BC} can be derived by using equation \eqref{eq:identity_1} and \eqref{eq:identity_2} in Appendix \ref{app:A}. For the $C_{2N}$ case, $\delta m$ in \eqref{eq:rh0BC} is an even number, which can be absorbed by $n$, so the structure of the solution to the integral \eqref{eq:integralEven} is similar to the $D_{2N}$ case. We have
\begin{align}\label{eq:BZl}
    Z_{C,2\ell}=\frac{(-1)^{\ell}2^{8\ell+1}}{(\theta_2\theta_3)^{8\ell}}\left(\sum_{n_j,n_j^{\prime}\in\mathbb{Z}}\frac{q^{2(n_1+\cdots n_{2\ell})+2\ell}\prod_{j=1}^{\ell}c(n_{2j-1}-n_{2j})c(n_{2j}^{\prime}-n_{2j+1}^{\prime})}{\prod_{j=1}^{2\ell}(1-q^{2n_{j}\pm 2n_{j}^{\prime}+1})}\right),
\end{align}
with the notation $n^{\prime}_{2\ell+1}=n^{\prime}_1$ and 
\begin{align}
    Z_{C,2\ell+1}=0.
\end{align}
Finally, the Schur index for $C_{2N}$ group is
\begin{align}
    \mathcal{I}_{C_{2N}}=\frac{1}{2^{2N}}\left(\frac{\eta^3}{\theta_4}\right)^{2N}\sum_{m_{\ell}}^{\prime}\prod_{\ell}\frac{Z_{C,2\ell}^{m_{\ell}}(-1)^{m_{\ell}}}{m_{\ell}!(2l)^{m_{\ell}}},
\end{align}
where the prime denotes a sum over non-negative integers $m_{\ell}$ that satisfy $\sum_{\ell=1}^{N}\ell m_{\ell}=N.$

Our method also applies to $B_{2N}$ case. However, since $\delta=1$, the shift $\delta m$ in \eqref{eq:rh0BC} changes the even/odd properties of $n$ in the summation. So the expression is more complicated. We experimentally test the calculation by expanding the $q$ series for the first few ranks, we find the agreement between the $B_{2N}$ and $C_{2N}$ Schur indices.

\subsubsection{Odd ranks}
In this subsection, we compute the Schur indices for $BCD$-groups with odd ranks. We will start with $D_{2N-1}$ case and then generalize it to $B_{2N-1}$ and $C_{2N-1}$ cases.
\paragraph{\underline{$D_{2N-1}$}}
According to Proposition \ref{Proposition1}, the determinant formula for $D_{2N-1}$ gauge group is represented with a $N\times N$ matrix where the entries of the matrix are not completely in the same pattern. To perform the integral with the help of the Fermi gas approach, we assume that there is an additional integral $\int_{0}^{2\pi \ri}d {a_{2N}}$, which does not change the result. Then if all the $a_{i}$ in the integral are not equal to $a_{2N}$, we have the integrals that were studied in the last subsection. If one of the $a_i$ in the integral is $a_{2N}$, let us suppose $a_{\ell}=a_{2N}$, we will encounter the integral in the form
\begin{equation}\label{eq:Ztilde}
   \widetilde{Z}_{\ell}= -\int_{0}^{2\pi \ri} d a_1 \cdots d{a_{\ell-1}}\rho(a_1,a_2) \cdots \rho(a_{\ell-2},a_{\ell-1}).
\end{equation}
The result of the integral \eqref{eq:Ztilde} is quite similar to the $D_{2N}$ case, we have
\begin{align}
    \widetilde{Z}_{2\ell+2}&=-\frac{2^{4\ell+1}}{(\theta_2\theta_3)^{4\ell}}\left(\sum_{\substack{n_{2\ell}=0\\ n_{j<2\ell}\in\mathbb{Z}}}\frac{q^{2(n_1+\cdots n_{2\ell-1})+\ell}}{\prod_{j=1}^{\ell}(1-q^{2n_{2j-1}\pm 2n_{2j}+1})(1-q^{2n_{2j}\pm (2n_{2j+1}+1)})}\right)\nonumber\\
    &=-2^{\ell+1}\left(\frac{\eta^3}{\theta_4}\right)^{-4\ell}\left(\sum_{\substack{n_{2\ell}=0\\ n_{j<2\ell}\in\mathbb{Z}}}\frac{q^{2(n_1+\cdots n_{2\ell})+\ell}\prod_{j=1}^{\ell}2^{1-\delta_{0,n_{2j}}}}{\prod_{j=1}^{\ell}(1-q^{2n_{2j-1}\pm 2n_{2j}+1})(1-q^{2n_{2j}\pm (2n_{2j+1}+1)})}\right),
\end{align}
with the notation $n_{2\ell+1}=n_1$, and
\begin{align}
    \widetilde{Z}_{2\ell+1}=0.
\end{align}
Then the Schur index of the gauge group $D_{2N-1}$ can be calculated by using a similar equation \eqref{eq:Sc_D2N}, but replacing one of the $Z_{2\ell}$ with $\widetilde{Z}_{2\ell}$ in the product. We have  
\begin{align}
    \mathcal{I}_{D_{2N-1}}=\frac{1}{2^{2N-2}}\left(\frac{\eta^3}{\theta_4}\right)^{2N-1}\sum_{m_{\ell}}^{\prime}\prod_{\ell}\frac{Z_{2\ell}^{m_{\ell}}(-1)^{m_{\ell}}}{m_{\ell}!(2\ell)^{m_{\ell}}}\sum_{k=1}^{\ell}{km_k}\frac{\widetilde{Z}_{2k}}{{Z}_{2k}},
\end{align}
where the prime denotes a sum over non-negative integers $m_{\ell}$ that satisfy $\sum_{\ell=1}^{N}\ell m_{\ell}=N.$

\paragraph{\underline{$B_{2N-1}$ and $C_{2N-1}$}}
The logic to calculate the $B_{2N-1}$ and $C_{2N-1}$ cases is similar to the $D_{2N-1}$. We assume that there is an additional integral $\int_{0}^{2\pi \ri}d {a_{2N}}$, which does not change the result. Then if all the $a_{i}$ in the integral are not equal to $a_{2N}$, we have the integrals that were studied in Section \ref{sec:Evenranks}. If one of the $a_i$ in the integral is $a_{2N}$, let us suppose $a_{\ell}=a_{2N}$, we will encounter the integral in the form
\begin{equation}
   \widetilde{Z}_{\ell}= -\int_{0}^{2\pi \ri} d a_1 \cdots d{a_{\ell-1}}\rho(a_1,a_2) \cdots \rho(a_{\ell-2},a_{\ell-1})\cdot \frac{\theta_1(\delta a_{1})^2}{\theta_4(\delta a_{1})^2}.
\end{equation}
Using equations \eqref{eq:rh0BC} and \eqref{eq:identity_2}, we determine that for the $C_{2N-1}$ case,
\begin{align}\label{eq:BZlpre}
    \widetilde{Z}_{C,2}=-4\mathcal{I}_{A_1},
\end{align}
if $\ell>0$,
\begin{align}
    &\widetilde{Z}_{C,2\ell+2}=\frac{(-1)^{\ell}2^{8\ell+1}}{(\theta_2\theta_3)^{8\ell}}\nonumber\\
    &\times\left(\sum_{ n_j,n_{j}^{\prime}\in\mathbb{Z}}\frac{q^{2(n_1+\cdots n_{2\ell})+2\ell}\prod_{j=1}^{\ell}c(n_{2j-1}-n_{2j})\cdot c(n^{\prime}_1)c(n^{\prime}_{2\ell})\prod_{j=2}^{\ell-1}c(n_{2j}^{\prime}-n_{2j+1}^{\prime})}{\prod_{j=1}^{2\ell}(1-q^{2n_{j}\pm 2n_{j}^{\prime}+1})}\right),
\end{align}
and 
\begin{align}
    \widetilde{Z}_{C,2\ell+1}=0.
\end{align}
Then the Schur index for $C_{2N-1}$ group is
\begin{align}
    \mathcal{I}_{C_{2N-1}}=\frac{1}{2^{2N-1}}\left(\frac{\eta^3}{\theta_4}\right)^{2N-1}\sum_{m_{\ell}}^{\prime}\prod_{\ell}\frac{Z_{C,2\ell}^{m_{\ell}}(-1)^{m_{\ell}}}{m_{\ell}!(2\ell)^{m_{\ell}}}\sum_{k=1}^{\ell}{km_k}\frac{\widetilde{Z}_{C,2k}}{{Z}_{C,2k}},
\end{align}
where the prime denotes a sum over non-negative integers $m_{\ell}$ that satisfy $\sum_{\ell=1}^{N}\ell m_{\ell}=N.$

Our method also applies to $B_{2N-1}$ case. However, similar to the $B_{2N}$ case, the expression is more complicated so we don't present it here. We experimentally test the calculation by expanding the $q$ series for the first few ranks, and we find the agreement between the $B_{2N-1}$ and $C_{2N-1}$ Schur indices.

\section{Fixing the exact modular formulas} \label{modular anomaly equations}

We shall try to fix the exact modular formulas for the $BCD$ types of gauge groups, using the calculations from the two methods in previous sections, as well as the modular properties in Table \ref{tab:modularform}. In this section we mostly use $\widetilde{\mathcal{I}}_G$ in (\ref{tildeI}), which has better modular formula than $\mathcal{I}_G$. The generators of the modular group $\Gamma(2)$ are 
$\theta_2(q)^{4},\theta_3(q)^{4}$. Due to the quasi-modularity, we also need to add the second Eisenstein series $E_2$. Similar to the $A_N$ case, the maximal weight can be easily read off from the formulas (\ref{eq:SI_def}, \ref{tildeI}) and grows linearly in the rank $N$. So the number of unknown coefficients of a generic quasi-modular ansatz goes like $N^3$ for large $N$, though the scaling factor here is larger since there is no universal symmetry between $\theta_2(q)^{4}$ and $\theta_3(q)^{4}$. Also similar to the $A_N$ case, the most significant constraint to fix the ansatz comes from the vanishing conditions that the $q$-series expansion starts from a very high power, scaling like $N^2$ for large rank $N$. 

For the $A_N$ case, there is a simple modular anomaly equation \cite{Huang:2022bry}, which fixes the dependence of $E_2$, so the number of unknown coefficients of the remaining modular ansatz goes like $N^2$, with a scaling factor smaller than that from the vanishing conditions. This enables the calculations of the Schur index for arbitrary rank in the $A_N$ cases. However, for the $BCD$ types of gauge groups, we do not find a simple modular anomaly equation, so the behaviors of the number of unknown coefficients $N^3$ is eventually bigger than the available conditions at a sufficiently large rank $N$.

For the $A_N$ cases, the quasi-modular formulas have contributions from each even weight up to the maximal weight, except for the absence of the constant term when $N$ is odd. For the $BCD$ types of gauge groups, there is an interesting new phenomenon that there is no term with weight smaller than one half of the maximal weight, so there is a non-trivial minimal modular weight. This constraint would reduce the number of unknown coefficients in the quasi-modular ansatz by about $\frac{1}{8}$. 

We observe an additional pattern that at each modular weight, the $q$-series expansion does not start from the generic constant term, but also from a high power at $q^{[\frac{N+1}{4}]}$ for the $SO(N)$ gauge group. If this is true, it would also provide additional conditions for helping to fix the quasi-modular ansatz at larger $N$. Furthermore, there is a pattern in the quasi-modular formulas that $\theta_2$ and $\theta_3$ are symmetric in the maximal weight terms, and further for the next maximal weight terms when $N$ is even. These constraints are nevertheless much less significant than the other conditions mentioned above. Of course, the results of $q$-series calculations from the previous two sections also provide additional conditions and redundant checks.

We summarize the conditions at the Table \ref{table1} and the formulas in Appendix \ref{results}. In the following, we discuss some more details for the $BC$-type and $D$-type gauge groups separately, and analyze some potential ansatz for modular anomaly equations.

\begin{table}
\begin{center}

	\begin{tabular} {|c|c|c|c|c||c|c|c|c||c|c|c|c|} 
		\hline  
			SO(N)  & 3 & 4 & 5 & 6 & 7 & 8 & 9 & 10& 11 & 12 & 13 & 14 \\
		 \hline 
			 maximal weight & 2 & 4 & 4 & 4 & 6 & 8 & 8 & 8& 10 & 12 & 12 & 12\\
		 \hline
		 	minimal weight & 2 & \underline{4} & 2 & 2 & 4 & 4 & 4 & 4& 6 & 6 & 6 & 6\\
		\hline
		leading order & $q$ & $q^2$ & $q^3$ & $q^4$ & $q^6$ & $q^8$ & $q^{10}$ & $q^{12}$ & $q^{15}$ & $q^{18}$ & $q^{21}$ & $q^{24}$\\
		\hline
	leading order of each weight& $q$ & $\underline{q^2}$ & $q$ & $q$ & $q^2$ & $q^2$ & $q^2$ & $q^2$ & $q^3$ & $q^3$ & $q^3$ & $q^3$\\
		\hline	 
		 
		\end{tabular}
	\caption{Some properties of the Schur indices for the $SO(N)$ group. There is a cyclic pattern when $N$ increases by 4. Because $SO(4)$ is not a simple Lie group, its behavior is an aberration of the usual patterns.}
\label{table1}
\end{center}

\end{table}

\subsection{$BC$-type gauge groups}

Counting the power of $q$ in the formulas (\ref{eq:SI_def}, \ref{tildeI}), we find the leading order in $q$-series expansion 
\be
	\widetilde{\mathcal{I}}_{B_N}(q)=\mathcal{O}(q^{\frac{N(N+1)}{2}}),
\ee
Using the conditions mentioned above, we fix some exact formulas in the following with more results in the Appendix \ref{results}
\be\ba
	\widetilde{\mathcal{I}}_{B_1}&=\widetilde{\mathcal{I}}_{A_1}= \frac{E_2}{2}+\frac{1}{24}\Theta_{0,1},\\
	\widetilde{\mathcal{I}}_{B_2}&=\frac{E_2^2}{4}+\frac{E_2}{24}\Theta_{0,1}
	+\frac{1}{48}\big(4\Theta_{0,2}-5\Theta_{1,1}+ 18\theta_2(q)^{4}\big),\\
	\widetilde{\mathcal{I}}_{B_3}&=\frac{E_2^3}{12}+\frac{E_2^2}{48}\Theta_{0,1}
	+\frac{E_2}{4608}\big( 8\Theta_{0,2}-\Theta_{1,1}+10\theta_2(q)^{4}+64\theta_3(q)^{4} \big),\\
	&+\frac{1}{82944}\big(4\Theta_{0,3}-15\Theta_{1,2}+ 15\theta_2(q)^{8}+96\theta_3(q)^{8}-33\theta_2(q)^{4}\theta_3(q)^{4} \big), 
\ea\ee
where 
	\be
		\Theta_{r,s}(q)=\theta_2(q)^{4r}\theta_2(q)^{4s}+\theta_2(q)^{4s}\theta_3(q)^{4r}.
	\ee
is the generator of $\Gamma^{0}(2)$ modular group. The case of $N=1,2$ can be obtained by using the formula in the literature \cite{Pan:2021mrw} to take the limit of the flavor parameter $b\rightarrow 0$.

Analogous to the $A_N$ case \cite{Huang:2022bry}, we test the various ansatzes for the modular anomaly equation with the available formulas. After some trials, we observed a simplest equation with the following form 
\be
	\partial_{E_2}\widetilde{\mathcal{I}}_{B_{N}}=\sum_{k=1}^N c^N_k \widetilde{\mathcal{I}}_{B_{N-k}},
\ee
where we use the convenient initial condition $\widetilde{\mathcal{I}}_{B_0}=1$. In contrast to the $A_N$ case, the coefficients $c^N_k$ are no longer constants, but dependent on $\theta_2^4$ and $\theta_3^4$. Additionally, for the case $c^N_2(E_2, \theta_2,\theta_3)$ we also need to introduce a $E_2$ dependence. This aberration has the compensating effect of simplifying the higher $c^N_k$ coefficients. The first few coefficients are 
\be\ba
	c^N_1(\theta_2,\theta_3)&=\frac{1}{2},\\
	c^N_2(\theta_2,\theta_3)&=\frac{1}{2} \widetilde{\mathcal{I}}_{B_1} =  \frac{E_2}{4}+\frac{1}{48}\Theta_{0,1} , \\
	c^N_3(\theta_2^4,\theta_3^4)&=c_1 \theta_2^4+c_2\theta_3^4, \\
	c^N_4(\theta_2,\theta_3)&= - \frac{\Theta_{1,1}}{1536}.
\ea\ee
We see that the coefficients $c^N_1, c^N_2, c^N_4$ are actually independent of $N$, while $c^N_3$ is a linear combination of $\theta_2^4$ and $\theta_3^4$ with coefficients dependent on $N$. For example, some specific anomaly equations are:
\be\ba
	\partial_{E_2}\widetilde{\mathcal{I}}_{B_2}&=\frac{1}{2}\widetilde{\mathcal{I}}_{B_1}+\frac{1}{2}\widetilde{\mathcal{I}}_{B_1} = \widetilde{\mathcal{I}}_{B_1},\\
	\partial_{E_2}\widetilde{\mathcal{I}}_{B_3}&=\frac{1}{2}\widetilde{\mathcal{I}}_{B_2}+\frac{1}{2}\widetilde{\mathcal{I}}_{B_1}^2+\frac{1}{576}(-\theta_2^4+8\theta_3^4),\\
	\partial_{E_2}\widetilde{\mathcal{I}}_{B_4}&=\frac{1}{2}\widetilde{\mathcal{I}}_{B_3}+\frac{1}{2}\widetilde{\mathcal{I}}_{B_2}\widetilde{\mathcal{I}}_{B_1}+\frac{1}{576}(23\theta_2^4-4\theta_3^4)\widetilde{\mathcal{I}}_{B_1}- \frac{\Theta_{1,1}}{1536}.
\ea\ee
For $k\geq 4$, the coefficient  $c^N_k$ is a polynomial of $ \theta_2^4$ and $ \theta_3^4$. Of course they are constrained by the maximal and minimal weights in $\widetilde{\mathcal{I}}_{B_{N}}$. We find that not all generic terms appear and many terms actually vanish. However, we do not otherwise identity any particularly simple pattern in these higher coefficients. Overall, although this is not as nice as the $A_N$ case, the modular anomaly is not completely random and still contains some useful information.

\subsection{$D$-type gauge groups} 
Counting the power of $q$ in the formulas (\ref{eq:SI_def}, \ref{tildeI}), we find the leading order in $q$-series expansion 
\be\ba
	\widetilde{\mathcal{I}}_{D_{2N}}&=\mathcal{O}(q^{2N^2}),\\
	\widetilde{\mathcal{I}}_{D_{2N+1}}&=\mathcal{O}(q^{2N(N+1)}),
\ea\ee
Despite the initial term of the $q$-expansion being an even number, the series also contains odd numbers. Again, using the conditions mentioned above, we fix some exact formulas in the following with more results in the Appendix \ref{results}
\be\ba
	\widetilde{\mathcal{I}}_{D_1}&=\widetilde{\mathcal{I}}_{B_1}=\widetilde{\mathcal{I}}_{A_1}= \frac{E_2}{2}+\frac{1}{24}\Theta_{0,1},\\
	\widetilde{\mathcal{I}}_{D_2}&=\widetilde{\mathcal{I}}_{A_1}^2=\frac{E_2^2}{4}+\frac{E_2}{24}\Theta_{0,1}+\frac{1}{576}(\Theta_{0,2}+\Theta_{1,1}),\\
	\widetilde{\mathcal{I}}_{D_3}&=\widetilde{\mathcal{I}}_{A_3} = \frac{E_2^2}{8}+\frac{E_2}{48}(2+\Theta_{0,1})+\frac{1}{1152}(\Theta_{0,2}-2\Theta_{1,1}+4\Theta_{0,1}),\\
	\widetilde{\mathcal{I}}_{D_4}&=\frac{E_2^4}{64}+\frac{E_2^3}{192}\Theta_{0,1}+\frac{E_2^2}{1536}(\Theta_{0,2}+\Theta_{1,1}+8\Theta_{0,1})\\
	&+\frac{E_2}{27648}(\Theta_{0,3}+3\Theta_{1,2}+24\Theta_{0,2}+6\Theta_{1,1})\\
	&+\frac{1}{1327104}(\Theta_{0,4}+4\Theta_{1,3}-24\Theta_{2,2}+48\Theta_{0,3}+162\Theta_2^8).
\ea\ee
Since the properties of the Schur indices for the $D_N$ group exhibit distinctions based on whether $N$ is even or odd, we treat them separately in our ansatzes for the modular anomaly equation. The simplest equation from our search is 
\be\ba
	\partial_{E_2}\widetilde{\mathcal{I}}_{D_{2N}}&=\sum_{k=1}^N \big(d_k^N \widetilde{\mathcal{I}}_{D_{2(N-k)}} \widetilde{\mathcal{I}}_{D_1}+e_k^N \widetilde{\mathcal{I}}_{D_{2(N-k)}}\big),\\
	\partial_{E_2}\widetilde{\mathcal{I}}_{D_{2N+1}}&=\sum_{k=1}^N \big(f_k^N \widetilde{\mathcal{I}}_{D_{2(N-k)+1}}  \widetilde{\mathcal{I}}_{D_1}+g_k^N \widetilde{\mathcal{I}}_{D_{2(N-k)+1}}\big),
\ea\ee
where we use the convention $\widetilde{\mathcal{I}}_{D_0}=1$. Because there is a weight difference of 4 between $\widetilde{\mathcal{I}}_{D_{2N}}$ and $\widetilde{\mathcal{I}}_{D_{2(N-1)}}$, we include the term $\widetilde{\mathcal{I}}_{D_N}\widetilde{\mathcal{I}}_{D_1}$. The low order coefficients $d,e,f,g$ are again somewhat simple 
\be\ba
	d_1^N&=f_1^N=\frac{1}{2},\\
	e_1^N&=0,~g_1^N=\frac{1}{24}, \\
	d_2^N&\sim \Theta_{0,1},\\
	f_2^N&=-\frac{1}{288}+\frac{1}{96}\Theta_{0,1}. 
\ea\ee
However, similar to the previous case, there seems no simple pattern in other higher coefficients as polynomials of $ \theta_2^4$ and $ \theta_3^4$ although many generic terms actually vanish.

\section{Discussions} \label{SecDiscuss}

We develop and improve some methods for calculating the unflavored Schur index in $\mathcal{N}=4$ super-Yang-Mills theory. The main results for $BCD$-type groups are listed in Appendix \ref{results} and the formula for the $G_2$ case is (\ref{formulaG2}). 

There are some remaining questions for potential future research. It would be certainly interesting to find more constraints which would enable the complete calculations of Schur indices for $BCD$-type groups of arbitrary rank, possibly from some improvements of the modular anomaly equations. 

It would be interesting to derive the various empirical features for the formulas in Appendix \ref{results}. In particular, the existence of a minimal modular weight, which is one half of the maximal weight, is an intriguing new feature which does not appear in the $A_N$ case, and seems deserving further study. Some mathematical techniques e.g. in the review paper \cite{Rastelli:2016tbz} may be useful to provide a proof of the non-trivial identity $\mathcal{I}_{B_N} = \mathcal{I}_{C_N}$ expected from S-duality.

The calculation for the $G_2$ case is simple due to the smallness of its rank. Further improvements in computational techniques are needed in order to fix the exact formulas for $F_4, E_{6,7,8}$ gauge groups, completing the picture for exceptional groups. 

For the $A_N$ case, the Schur index is related to the generalized MacMahon's sum-of-divisors functions \cite{fcd25745b70a45868e37b97d86434d65, cite-key, Kang:2021lic}. However, the congruence subgroups in \cite{cite-key} do not match those of the Schur indices of $BCD$-type groups. It would be interesting to search for more generalizations of such MacMahon's sum-of-divisors functions that could give the Schur indices in this paper.

\vspace{0.2in} {\leftline {\bf Acknowledgments}}

We thank Sheldon Katz, Albrecht Klemm for stimulating collaborations on related papers and Jun-Hao Li, Gao-fu Ren, Pei-xuan Zeng for helpful discussions. XW thanks Tadashi Okazaki and Yongchao L\"u for helpful discussions. BD and MH thank Chiung Hwang, Sung-Soo Kim, Yiwen Pan, Futoshi Yagi, Wenbin Yan for related enlightening lectures at the PCFT (Peng Huanwu Center for Fundamental Theory) mini advanced school ``Supersymmetric field theories and related topics" in August 2023. The works of MH was supported in parts by National Natural Science Foundation of China (Grant No. 12247103). XW is supported by a KIAS Individual Grant QP079201.
\appendix

\section{Elliptic functions and modular forms}\label{app:A}
In this appendix, we summarize some basic definitions for various elliptic functions that we have used in the main text.

\paragraph{Jacobi theta functions}
The Jacobi theta functions $\theta_i(z;\tau),i=1,\cdots,4,$ are defined as follows:
\begin{align}
    \theta_1(z;\tau)&=\ri\sum_{n\in\mathbb{Z}} (-1)^n q^{(n+\frac{1}{2})^2}e^{ (n+\frac{1}{2})z},\\
    \theta_2(z;\tau)&=\sum_{n\in\mathbb{Z}}  q^{(n+\frac{1}{2})^2}e^{ (n+\frac{1}{2})z},\\
    \theta_3(z;\tau)&=\sum_{n\in\mathbb{Z}}  q^{n^2}e^{ n z },\\
    \theta_4(z;\tau)&=\sum_{n\in\mathbb{Z}} (-1)^n q^{\frac{1}{2}n^2}e^{nz},
\end{align}
where $q=e^{ \pi \ri \tau}$. We denote $\theta_i=\theta_i(0;\tau),i=2,3,4$ as theta constants.
\paragraph{Dedekind eta function}
The Dedekind eta function is defined by
\begin{equation}
    \eta(\tau)=q^{\frac{1}{12}}\prod_{n=1}^{\infty}(1-q^{2n}).
\end{equation}
It is related to the theta constants by
\begin{equation}
    \eta(\tau)^3=\frac{1}{2}\theta_2\theta_3\theta_4.
\end{equation}
If there is no confusion regarding notations, we also use $\eta$ to denote the Dedekind eta function.
\paragraph{Jacobi elliptic functions}
The Jacobi elliptic functions are a set of basic elliptic functions. In this section, we will focus on the elliptic sine function $\mathrm{sn}(z,k)$, and review some of its properties which can also be found in the textbook \cite{MR1007595}. The elliptic sine function can be defined from the Jacobi theta functions as \footnote{Here we don't use the standard notation, in order to avoid unnecessary additional definitions.}:
\begin{equation}
    \mathrm{sn}(z\theta_3^2)=\frac{\theta_3}{\theta_2}\frac{\theta_1(z)}{\theta_4(z)}
    =\frac{2\ri}{\theta_2^2}\sum_{n\in\mathbb{Z}}\frac{q^{n+\frac{1}{2}}e^{(n+\frac{1}{2})z}}{1-q^{2n+1}},
\end{equation}
so we can derive the expansion
\begin{equation}\label{eq:identity_1}
    \frac{\theta_1(z)}{\theta_4(z)}=\frac{2\ri}{\theta_2\theta_3}\sum_{n\in\mathbb{Z}}\frac{q^{n+\frac{1}{2}}e^{(n+\frac{1}{2})z}}{1-q^{2n+1}}.
\end{equation}
Similarly, we can derive the expansion
\begin{align}\label{eq:identity_2}
    \frac{\theta_1(z)^2}{\theta_4(z)^2}=-\frac{4}{{\theta_2^2\theta_3^2}}\left(-2\mathcal{I}_{A_1}+\sum_{n=1}^{\infty}\frac{n q^{n}}{(1-q^{2n})}(e^{n z  }+e^{-n z  })\right),
\end{align}
where $\mathcal{I}_{A_1}$ is the $\mathcal{N}=4$ Schur index for $SU(2)$ gauge group, it has the expression
\begin{align}
    \mathcal{I}_{A_1}=\mathcal{I}_{SU(2)}=\sum_{n=1}^{\infty}\frac{q^{2n-1}}{(1-q^{2n-1})^2}={\frac{1}{2}E_2+\frac{1}{24}(\theta_2^4+\theta_3^4)}.
\end{align}

\section{Ring of modular forms under congruence subgroups of $\mathrm{SL}_2(\mathbb{Z})$} \label{Seccongruence}

We summarize some known facts about modular forms of congruence subgroups. Some recent references are \cite{MR2385372,Dabholkar:2012nd, Cota:2019cjx}.

\paragraph{Modular form of $\mathrm{SL}_2(\mathbb{Z})$}
A modular form of weight $k$ is a holomorphic function $f: \mathcal{H}\rightarrow \mathbb{C}$ on the upper-half plane that satisfying:
\begin{align}
    f\left(\frac{a\tau+b}{c\tau+d}\right)=(c\tau+d)^k f(\tau),
\end{align}
for all $\begin{pmatrix}
    a&b\\
    c&d
\end{pmatrix}\in\mathrm{SL}_2(\mathbb{Z})$. The modular group $\mathrm{SL}_2(\mathbb{Z}) $ is defined as
\begin{align}
    \mathrm{SL}_2(\mathbb{Z}) := \left\{\begin{pmatrix}
    a&b\\
    c&d
\end{pmatrix}\,\middle|\, a,b,c,d\in\mathbb{Z},ad-bc=1\right\}.
\end{align}
Define the Eisenstein series 
\begin{align}
    E_{2k}(\tau)=-\frac{B_{2k}}{(2k)!}\left(1+\frac{4k}{B_{2k}}\sum_{n=1}^{\infty} \frac{n^{2k-1}q^{2n}}{1-q^{2n}}\right),\quad k>0
\end{align}
where $q=e^{\ri \pi \tau}$ and $B_{2k}$ is Bernoulli number with the values $B_2=\frac{1}{6},B_4=-\frac{1}{30},B_6=\frac{1}{42},\cdots,$ for $k=1,2,3,\cdots$. Then the even weight modular forms $M_{*}(\mathrm{SL}_2(\mathbb{Z}),\tau)=\bigoplus\limits_{k=0}^{\infty}M_{2k}(\mathrm{SL}_2(\mathbb{Z}),\tau)$ of $\mathrm{SL}_2(\mathbb{Z})$ are finitely generated by the fourth and sixth Eisenstein series
\begin{equation}
    M_{*}(\mathrm{SL}_2(\mathbb{Z}),\tau)=\mathbb{C}[E_4(\tau),E_6(\tau)].
\end{equation}
The second Eisenstein series $E_2(\tau)$ is not a modular form; however, it admits a non-holomorphic completion
\begin{equation}
    \widehat{E}_2(\tau,\bar{\tau})=E_2(\tau)+\frac{1}{4\pi\, \mathrm{Im}\tau},
\end{equation}
which transforms as a weight-two modular form. We call $\widehat{E}_2(\tau,\bar{\tau})$ the almost-holomorphic modular form, and the holomorphic part $E_2(\tau)$ is called the quasi-modular form. The ring of quasi-modular form for a congruence subgroup is consistent of the ring of modular form together with $E_2(\tau)$.
\paragraph{Modular form of congruence subgroups}
A modular form of weight $k$, level $n$ is a holomorphic function $f: \mathcal{H}\rightarrow \mathbb{C}$ satisfying:
\begin{align}
    f\left(\frac{a\tau+b}{c\tau+d}\right)=(c\tau+d)^k f(\tau),
\end{align}
for all $\begin{pmatrix}
    a&b\\
    c&d
\end{pmatrix}\in \Gamma_0(n)$. The subgroup $\Gamma_0(n)\in\mathrm{SL}_2(\mathbb{Z})$, which are called the Hecke congruence subgroup of level $n$, is defined as
\begin{align}
    \Gamma_0(n) := \left\{\begin{pmatrix}
    a&b\\
    c&d
\end{pmatrix}\in \mathrm{SL}_2(\mathbb{Z}):c\equiv 0 \quad(\mathrm{mod}\,\,n)\right\}.
\end{align}
\begin{align}
    \Gamma^0(n) := \left\{\begin{pmatrix}
    a&b\\
    c&d
\end{pmatrix}\in \mathrm{SL}_2(\mathbb{Z}):b\equiv 0 \quad(\mathrm{mod}\,\,n)\right\}.
\end{align}
We also encounter the intersection of the congruence subgroups
\begin{align}
    \Gamma_0(n_1)\cap\Gamma^0(n_2) := \left\{\begin{pmatrix}
    a&b\\
    c&d
\end{pmatrix}\in \mathrm{SL}_2(\mathbb{Z}):c\equiv 0 \quad(\mathrm{mod}\,\,n_1) \,\, \& \,\,  b\equiv 0 \quad(\mathrm{mod}\,\,n_2)\right\}.
\end{align}
It is clear to see, if $f(\tau)$ is a modular form of $\Gamma_0(n_1n_2)$, then $f(\tau/n_2)$ is a modular form of $\Gamma_0(n_1)\cap\Gamma^0(n_2)$.
Define 
\begin{equation}\label{eq:E3}
    E_2^{(n)}(\tau)=-\frac{24}{n-1}q\partial_q \log \frac{\eta(\tau)}{\eta(n\tau)}, 
\end{equation}
the rings of the even weight modular forms $M_{*}(\Gamma,\tau)=\bigoplus\limits_{k=0}^{\infty}M_{2k}(\Gamma,\tau)$ provided in SageMath \cite{sagemath} can be expressed as
\begin{align}
    &M_{*}(\Gamma_0(2),\tau)=\mathbb{C}[E_2^{(2)}(\tau),E_4(\tau)],\\
    &M_*(\Gamma_0(3),\tau)=\mathbb{C}[E_2^{(3)}(\tau),E_4(\tau),E_6(\tau)],\\
    &M_*(\Gamma_0(4),\tau)=\mathbb{C}[E_2^{(2)}(\tau),E_2^{(4)}(\tau)],\\
    &M_*(\Gamma_0(6),\tau)=\mathbb{C}[E_2^{(2)}(\tau),E_2^{(3)}(\tau),E_2^{(6)}(\tau)],\\
    &M_*(\Gamma_0(12),\tau)=\mathbb{C}[E_2^{(2)}(\tau),E_2^{(3)}(\tau),E_2^{(4)}(\tau),E_2^{(6)}(\tau),E_2^{(12)}(\tau)],
\end{align}
and
\begin{equation}
    M_*(\Gamma_0(n)\cap \Gamma^0(2),\tau)=M_*(\Gamma_0(2n),\frac{1}{2}\tau).
\end{equation}

\section{Proof of Proposition 1}\label{app:proof}
Denote the functions on the left-hand side and on the right-hand side of equation \eqref{eq:cauchyDv2} as $f_{\mathrm{L}}(a_1,\cdots,a_{2N})$ and $f_{\mathrm{R}}(a_1,\cdots,a_{2N})$ respectively. 
It is easy to see that $f_{\mathrm{L}}(a_1,\cdots,a_{2N})$ and $f_{\mathrm{R}}(a_1,\cdots,a_{2N})$ are elliptic functions with the same arguments. The denominator of $f_{\mathrm{L}}(a_1,\cdots,a_{2N})$ is divisible by the denominator of each term in the determinant expression of $f_{\mathrm{R}}(a_1,\cdots,a_{2N})$. So the poles of $f_{\mathrm{L}}(a_1,\cdots,a_{2N})$ contain those of $f_{\mathrm{R}}(a_1,\cdots,a_{2N})$ including multiplicity. We note that the converse is not obviously true since the poles in $f_{\mathrm{R}}(a_1,\cdots,a_{2N})$ may cancel among different terms in the determinant. In the followings we also prove that the zero points of $f_{\mathrm{R}}$ contain those of $f_{\mathrm{L}}$ including multiplicity, so $f_{\mathrm{R}}/f_{\mathrm{L}}$ would be an analytic function with no pole. According to Liouville's theorem on elliptic functions,  it would be  just a constant. 

It is easy to see that all the zero points of $f_{\mathrm{L}}(a_1,\cdots,a_{2N})$ are of degree two and they are located at the points
\begin{align}\label{eq:zeropoints}
    a_i=\pm a_j,\quad j\neq i.
\end{align}
Now we show that the $f_{\mathrm{R}}$ has the same zero points which are at least of degree two. Denote the entry of the matrix in the determinant to be
\begin{align}
    \rho(a_i,a_j)\equiv\frac{\theta_1(a_j\pm a_i)}{\theta_4(a_j \pm a_i)},
\end{align}
then $f_{\mathrm{R}}$ can be expanded as
\begin{align}\label{eq:C4}
    f_{\mathrm{R}}(a_1,\cdots,a_{2N})=\sum_{i_1,\cdots,i_{2N}}\varepsilon_{i_1\cdots i_{2N}}\rho(a_1,a_{i_1})\rho(a_2,a_{i_2})\cdots \rho(a_{2N},a_{i_{2N}}),
\end{align}
where $\varepsilon_{i_1\cdots i_{2N}}$ is the Levi-Civita symbol.
If $a_1=a_2$, then \eqref{eq:C4} is zero due to the antisymmetric summation. If we exchange $a_1$ and $a_2$, the determinant is obviously invariant. So the zero at $a_1=a_2$ must be of even degree, implying it is at least of degree two. By performing the reflection action $a_2\rightarrow -a_2$ and permutation action on the variables $a_j$, we find that all the zero points in \eqref{eq:zeropoints} are also at least degree two zero points of $f_{\mathrm{R}}$. We can conclude $f_{\mathrm{R}}/f_{\mathrm{L}}$ is analytic and thus is a constant. By using the asymptotic behavior of $f_{\mathrm{R}}/f_{\mathrm{L}}$, we find $f_{\mathrm{R}}/f_{\mathrm{L}}=1$, which finish the proof of equation \eqref{eq:cauchyDv2}. The proof of equation \eqref{eq:cauchyDv2_odd} can be done in the same manner that we have done.

In contrast to the $A_N$ case, we can not simply change the numerator of $\rho(a_i,a_j)$ to $\theta_{2,3}(a_i\pm a_j)$. If we do this, it would give additional poles $\theta_4(2a_i)=0$ from the diagonal matrix elements so that $f_R/f_L$ may not be analytic.

\section{Schur Indices for $SO(N)$ Groups up to $N=17$} \label{results}

This appendix presents the Schur indices $\widetilde{I}_{SO(N)}$ for the $SO(N)$ groups up to $N=17$. This includes the $B$ and $D$ types of gauge groups. The results for the $C$ type symplectic gauge groups are the same as those of the $B$ types with the same ranks. Some sporadic results on low ranks have appeared in the literature, e.g. \cite{Pan:2021mrw, Guo:2023mkn}. To show the symmetry of $\theta_2$ and $\theta_3$ near the maximal weight,  we also use the notation $\Theta_{i,j} \equiv \theta_2^{4i} \theta_3^{4j} +  \theta_3^{4i} \theta_2^{4j}$.

\be\ba
	\widetilde{\mathcal{I}}_{SO(3)}&= \widetilde{\mathcal{I}}_{SU(2)} = \frac{E_2}{2}+\frac{1}{24}\Theta_{0,1},\\
	\widetilde{\mathcal{I}}_{SO(4)}&= ( \widetilde{\mathcal{I}}_{SU(2)})^2  = \frac{E_2^2}{4}+\frac{E_2}{24}\Theta_{0,1}+\frac{1}{576}(\Theta_{0,2}+\Theta_{1,1}),\\
	\widetilde{\mathcal{I}}_{SO(5)}&=\frac{E_2^2}{4}+\frac{E_2}{24}\Theta_{0,1}
	+\frac{1}{1152}\big(2\Theta_{0,2}-5\theta_2^4\theta_3^4+ 9\theta_2^{4}\big),\\
	\widetilde{\mathcal{I}}_{SO(6)}&= \widetilde{\mathcal{I}}_{SU(4)}  = \frac{E_2^2}{8}+\frac{E_2}{48}(2+\Theta_{0,1})+\frac{1}{1152}(\Theta_{0,2}-2\Theta_{1,1}+4\Theta_{0,1}),\\
	\widetilde{\mathcal{I}}_{SO(7)}&=\frac{E_2^3}{12}+\frac{E_2^2}{48}\Theta_{0,1}
	+\frac{E_2}{4608}\big( 8\Theta_{0,2}-\Theta_{1,1}+10\theta_2^{4}+64\theta_3^{4} \big)\\
	&+\frac{1}{82944}\big(4\Theta_{0,3}-15\Theta_{1,2}+ 15\theta_2^{8}+96\theta_3^{8}-33\theta_2^{4}\theta_3^{4} \big),\\ 
	\widetilde{\mathcal{I}}_{SO(8)}&=\frac{E_2^4}{64}+\frac{E_2^3}{192}\Theta_{0,1}+\frac{E_2^2}{1536}(\Theta_{0,2}+\Theta_{1,1}+8\Theta_{0,1})\\
	&+\frac{E_2}{27648}(\Theta_{0,3}+3\Theta_{1,2}+24\Theta_{0,2}+6\Theta_{1,1})\\
	&+\frac{1}{1327104}(\Theta_{0,4}+4\Theta_{1,3}-24\Theta_{2,2}+48\Theta_{0,3}+162\theta_2^8), 
	\ea\ee
\be\ba		
	\widetilde{\mathcal{I}}_{SO(9)}&=\frac{5E_2^4}{192}+\frac{5E_2^3}{576}\Theta_{0,1}+\frac{E_2^2}{9216}(10\Theta_{0,2}+\Theta_{1,1}+106\theta_2^4+16\theta_3^4)\\
	&+\frac{E_2}{82944}(5\Theta_{0,3}-12\Theta_{1,2}+159\theta_2^8+24\theta_3^8+3\theta_2^4\theta_3^4)\\
	&+\frac{1}{15925248}(20\Theta_{0,4}-136\Theta_{1,3}+87\Theta_{2,2}+1272\theta_2^{12}\\	&+192\theta_3^{12}-1116\theta_2^8\theta_3^4-1224\theta_2^4\theta_3^8+4374\theta_2^8), \\	
	\widetilde{\mathcal{I}}_{SO(10)}&=\frac{E_2^4}{128}+\frac{ E_2^3}{384}(2+\Theta_{0,1} )+\frac{E_2^2} {3072}(12\Theta_{0,1}+\Theta_{0,2})+\\	
&\frac{E_2}{276480}
(5\Theta_{0,3}-15\Theta_{1,2}+150\Theta_{0,2}-30\Theta_{1,1}+36\theta_2^4+576\theta_3^4)\\
&+\frac{1}{13271040}(5\Theta_{0,4}-40\Theta_{1,3}+36\Theta_{2,2}+280\Theta_{0,3}-600\Theta_{1,2}-504\Theta_{1,1}+954\theta_2^8+2304 \theta_3^8), \\
\widetilde{\mathcal{I}}_{SO(11)}&=\frac{13E_2^5}{1920}+\frac{13E_2^4}{4608}\Theta_{0,1}+\frac{E_2^3}{27648}(13\Theta_{0,2}+4\Theta_{1,1}+6\theta_2^4+168\theta_3^4)\\
	&+\frac{E_2^2}{331776}(13\Theta_{0,3}-15\Theta_{1,2}+18\theta_2^8+504\theta_3^8+162\theta_2^4\theta_3^4)\\
	&+\frac{E_2}{398 131 200}(650\Theta_{0,4}-2800\Theta_{1,3}-825\theta_2^8\theta_3^8+1800\theta_2^{12}\\	&+50400\theta_3^{12}-24750\theta_2^8\theta_3^4-12600\theta_2^4\theta_3^8  +459\theta_2^8+221 184\theta_3^8+63936\theta_2^4\theta_3^4)\\
	&+\frac{1}{4777574400}( 130\Theta_{0,5}-1150\Theta_{1,4}+1975\Theta_{2,3}+600\theta_2^{16}+16800\theta_3^{16}\\
&-24150\theta_2^{12}\theta_3^4-28200\theta_2^4\theta_3^{12}-4950\theta_2^8\theta_3^8+459\theta_2^{12}+221184\theta_3^{12} -72981\theta_2^8\theta_3^4-46656\theta_2^4\theta_3^8), \\
\widetilde{\mathcal{I}}_{SO(12)}&=\frac{E_2^6}{1536}+\frac{ E_2^5}{3072}\Theta_{0,1}+\frac{E_2^4} {73728}(48\Theta_{0,1}+5\Theta_{0,2}+5\Theta_{1,1})\\	
&+\frac{E_2^3}{663552}
(5\Theta_{0,3}+15\Theta_{1,2}+144\Theta_{0,2}+90\Theta_{1,1})\\
&+\frac{E_2^2}{53084160}(25\Theta_{0,4}+100\Theta_{1,3}-60\Theta_{2,2}+1440\Theta_{0,3}\\
&+2160\Theta_{1,2}+4224\Theta_{1,1}+16842\theta_2^8+3072 \theta_3^8)\\
&+\frac{E_2}{318 504 960} ( 5\Theta_{0,5}+25\Theta_{1,4}-220\Theta_{2,3}+480\Theta_{0,4}+840\Theta_{1,3}\\
&-720\Theta_{2,2}+16842\theta_2^{12}+3072\theta_3^{12}+1242 \theta_2^8 \theta_3^4+6912 \theta_2^4 \theta_3^8)\\
&+\frac{1}{22 932 357 120} ( 5\Theta_{0,6}+30\Theta_{1,5}-735\Theta_{2,4}+1184\Theta_{3,3}+720\Theta_{0,5}+1440\Theta_{1,4}-12240\Theta_{2,3}\\
&+50526\theta_2^{16}+9216\theta_3^{16}-17892\theta_2^{12}\theta_3^4+16128\theta_2^{4}\theta_3^{12}-40770\theta_2^8\theta_3^8+174960\theta_2^{12}),
\ea\ee
\be\ba
	\widetilde{\mathcal{I}}_{SO(13)}&=\frac{19E_2^6}{11520}+\frac{19E_2^5}{23040}\Theta_{0,1}+\frac{E_2^4}{221184}(38\Theta_{0,2}+31\theta_2^4\theta_3^4+589\theta_2^4-32\theta_3^4)\\
	&+\frac{E_2^3}{1990656}(38\Theta_{0,3}-21\Theta_{1,2}+1767\theta_2^8-96\theta_3^8+735\theta_2^4\theta_3^4)\\
	&+\frac{E_2^2}{796 262 400}(950\Theta_{0,4}-2950\Theta_{1,3}-1725\theta_2^8\theta_3^8+88350\theta_2^{12}\\	&-4800\theta_3^{12}+6750\theta_2^8\theta_3^4-25650\theta_2^4\theta_3^8  +435099\theta_2^8-24576\theta_3^8+162096 \theta_2^4\theta_3^4)\\
	&+\frac{E_2}{4777574400}( 190\Theta_{0,5}-1300\Theta_{1,4}+1225\Theta_{2,3}+29450\theta_2^{16}-1600\theta_3^{16}\\
&-31600\theta_2^{12}\theta_3^4-32950\theta_2^4\theta_3^{12}-11400 \theta_2^8\theta_3^8+435099\theta_2^{12}-24576\theta_3^{12} +45027 \theta_2^8\theta_3^4-74448\theta_2^4\theta_3^8)\\
&+\frac{1}{687970713600}(  380\Theta_{0,6}-4470\Theta_{1,5}+15150\Theta_{2,4}-14675\Theta_{3,3}+88350\theta_2^{20}-4800 \theta_3^{20}\\
&-267900\theta_2^{16}\theta_3^4-182850 \theta_2^4\theta_3^{16}+161925\theta_2^{12}\theta_3^8+176100 \theta_2^8\theta_3^{12}+
2610594\theta_2^{16}-147456\theta_3^{16}\\
&-1908477\theta_2^{12}\theta_3^4-1865952 \theta_2^4\theta_3^{12}-614142 \theta_2^8\theta_3^8+8365275\theta_2^{12} ), \\
\widetilde{\mathcal{I}}_{SO(14)}&=\frac{E_2^6}{3072}+\frac{ E_2^5}{6144}(2+\Theta_{0,1})+\frac{E_2^4} {147456}(68\Theta_{0,1}+5\Theta_{0,2}+2\Theta_{1,1})+\\	
&\frac{E_2^3}{6635520}
(25\Theta_{0,3}-15\Theta_{1,2}+870\Theta_{0,2}+330\Theta_{1,1}-3492 \theta_2^4+6768\theta_3^4)\\
&+\frac{E_2^2}{106168320}(25\Theta_{0,4}-80\Theta_{1,3}-24\Theta_{2,2}+1640\Theta_{0,3}\\
&-840\Theta_{1,2}+6168\Theta_{1,1}+2874\theta_2^8+30144 \theta_3^8)\\
&+\frac{E_2}{4459069440} ( 35\Theta_{0,5}-245\Theta_{1,4}+224\Theta_{2,3}+3710\Theta_{0,4}-10360\Theta_{1,3}-840\Theta_{2,2}+69006 \theta_2^{12}\\
&+116256\theta_3^{12}-20664 \theta_2^4 \theta_3^8-67914 \theta_2^8 \theta_3^4-60372 \theta_2^8+442368 \theta_3^8+145152 \theta_2^4 \theta_3^4),\\
&+\frac{1}{321052999680} ( 35\Theta_{0,6}-420\Theta_{1,5}+1407\Theta_{2,4}-568\Theta_{3,3}+5460\Theta_{0,5}-33180 \Theta_{1,4}\\
&+31920 \Theta_{2,3}+255906 \theta_2^{16}+254016\theta_3^{16}-579096\theta_2^{12}\theta_3^4
-480816\theta_2^{4}\theta_3^{12}-16254 \theta_2^8\theta_3^8\\
&+862488\theta_2^{12}+2654208 \theta_3^{12}-1295352 \theta_2^8 \theta_3^4-456192 \theta_2^4 \theta_3^8),
\ea\ee
\be\ba
\widetilde{\mathcal{I}}_{SO(15)}&=\frac{29E_2^7}{80640}+\frac{29E_2^6}{138240}\Theta_{0,1}+\frac{E_2^5}{2211840}(116\Theta_{0,2}+115\theta_2^4\theta_3^4-367\theta_2^4+2144\theta_3^4)\\
	&+\frac{E_2^4}{15925248}(116\Theta_{0,3}-3\Theta_{1,2}-1101\theta_2^8+6432\theta_3^8+2883\theta_2^4\theta_3^4)\\
	&+\frac{E_2^3}{2388787200}	
(1450\Theta_{0,4}-2975\Theta_{1,3}-2775\theta_2^8\theta_3^8-27525 \theta_2^{12}\\	&+160800\theta_3^{12}-20700 \theta_2^8\theta_3^4+58275\theta_2^4\theta_3^8  -163089\theta_2^8+827136\theta_3^8+318744\theta_2^4\theta_3^4)\\
	&+\frac{E_2^2}{9555148800}( 290\Theta_{0,5}-1475\Theta_{1,4}+200\Theta_{2,3}-9175\theta_2^{16}+53600\theta_3^{16}\\
&-39175\theta_2^{12}\theta_3^4-22975\theta_2^4\theta_3^{12}-25575\theta_2^8\theta_3^8-163089\theta_2^{12}+827136\theta_3^{12}\\
&-29313\theta_2^8\theta_3^4+329112 \theta_2^4\theta_3^8)\\
&+\frac{E_2}{67421129932800}(56840\Theta_{0,6}-518910 \Theta_{1,5}+984900\Theta_{2,4}+442225\theta_2^{12}\theta_3^{12}\\
&-2697450\theta_2^{20}+15758400\theta_3^{20}-21403200\theta_2^{16}\theta_3^4-30333450\theta_2^4\theta_3^{16}+3156825\theta_2^{12}\theta_3^8\\
&-117600 \theta_2^8\theta_3^{12}
-95896332 \theta_2^{16}+486355968\theta_3^{16}-222500817\theta_2^{12}\theta_3^4\\
&-93012192\theta_2^4\theta_3^{12}-141703884\theta_2^8\theta_3^8-109024137\theta_2^{12}\\
& +1719926784 
\theta_3^{12}+236475936 \theta_2^8\theta_3^4+555393024\theta_2^4\theta_3^8)\\
&+\frac{1}{809053559193600}(  8120\Theta_{0,7}-115150\Theta_{1,6}+501270 \Theta_{2,5}\\
&-542675\Theta_{3,4}-539490 \theta_2^{24}+3151680 \theta_3^{24}-6337170 \theta_2^{20} \theta_3^4\\
&-13005090 \theta_2^4 \theta_3^{20}+13858425 \theta_2^{16} \theta_3^8+13560750 \theta_2^8 \theta_3^{16}+6508425 \theta_2^{12} \theta_3^{12}\\
&-31965444\theta_2^{20}+162118656\theta_3^{20}-142790949\theta_2^{16}\theta_3^4-224056224\theta_2^4\theta_3^{16}+8038107 \theta_2^{12}\theta_3^8\\
&-10120068 \theta_2^8\theta_3^{12}
-109024137 \theta_2^{16}+1719926784\theta_3^{16}-492641433\theta_2^{12}\theta_3^4\\
&-304570368\theta_2^4\theta_3^{12}-363706848\theta_2^8\theta_3^8  ), 
\ea\ee

\be\ba
\widetilde{\mathcal{I}}_{SO(16)}&=\frac{E_2^8}{49152}+\frac{ E_2^7}{73728}\Theta_{0,1}+\frac{E_2^6} {1769472}(72\Theta_{0,1}+7\Theta_{0,2}+7\Theta_{1,1})\\	
&+\frac{E_2^5}{10616832}
(7\Theta_{0,3}+21\Theta_{1,2}+216\Theta_{0,2}+162\Theta_{1,1})\\
&+\frac{E_2^4}{2548039680}(175\Theta_{0,4}+700\Theta_{1,3}+120\Theta_{2,2}+10800\Theta_{0,3}\\
&+21600\Theta_{1,2}+85248\theta_2^4\theta_3^4+219582\theta_2^8-67968 \theta_3^8)\\
&+\frac{E_2^3}{7644119040} ( 35\Theta_{0,5}+175\Theta_{1,4}-460\Theta_{2,3}+3600\Theta_{0,4}+9000\Theta_{1,3}\\
&+2160\Theta_{2,2}+219582 \theta_2^{12}-67968 \theta_3^{12}+133056 \theta_2^4 \theta_3^8+51246 \theta_2^8 \theta_3^4),\\
&+\frac{E_2^2}{321052999680} ( 245\Theta_{0,6}+1470\Theta_{1,5}-13335\Theta_{2,4}-1904\Theta_{3,3}+37800\Theta_{0,5}+113400\Theta_{1,4}\\
&-257040 \Theta_{2,3}+4611222 \theta_2^{16}-1427328\theta_3^{16}+362124\theta_2^{12}\theta_3^4
+3798144\theta_2^{4}\theta_3^{12}+1121526 \theta_2^8\theta_3^8\\
&+24867648\theta_2^{12}-3981312 \theta_3^{12}+10734768 \theta_2^8 \theta_3^4-+5971968\theta_2^4 \theta_3^8),\\
&+\frac{E_2}{7705271992320} ( 35\Theta_{0,7}+245\Theta_{1,6}-4935\Theta_{2,5}+11431\Theta_{3,4}\\
&+7560\Theta_{0,6}+26460 \Theta_{1,5}-234360\Theta_{2,4}+24192\Theta_{3,3}\\
&+1537074 \theta_2^{20}-475776 \theta_3^{20}-117306 \theta_2^{16} \theta_3^4\\
&+1600704 \theta_2^4 \theta_3^{16}-2311722 \theta_2^{12} \theta_3^8-1030302 \theta_2^8 \theta_3^{12}\\
&+24867648 \theta_2^{16}-3981312\theta_3^{16}+138024\theta_2^{12}\theta_3^4
+7962624\theta_2^{4}\theta_3^{12}+2150064  \theta_2^8\theta_3^8)\\
&\frac{1}{739706111262720}( 35\Theta_{0,8}+280\Theta_{1,7}-10360\Theta_{2,6}+65464\Theta_{3,5}-136546 \Theta_{4,4}\\
&+10080 \Theta_{0,7}+40320\Theta_{1,6}-846720\Theta_{2,5}+1574496\Theta_{3,4}\\
&+3074148 \theta_2^{24}-951552\theta_3^{24}-710640 \theta_2^{20} \theta_3^4+3870720 \theta_2^4 \theta_3^{20}-16414272 \theta_2^{16} \theta_3^8\\
&-12932892\theta_2^8 \theta_3^{16}+2805264\theta_2^{12} \theta_3^{12}+99470592 \theta_2^{20}-15925248 \theta_3^{20}-41834880 \theta_2^{16} \theta_3^4\\
&+39813120 \theta_2^4 \theta_3^{16}-69999552 \theta_2^{12} \theta_3^8-29719872\theta_2^8 \theta_3^{12}+322407540 \theta_2^{16}), \\
\ea\ee

\be\ba
	\widetilde{\mathcal{I}}_{SO(17)}&=\frac{191E_2^8}{2580480}+\frac{191E_2^7}{3870720}\Theta_{0,1}+\frac{E_2^6}{13271040}(191\Theta_{0,2}+211\theta_2^4\theta_3^4+4191\theta_2^4-1128\theta_3^4)\\
	&+\frac{E_2^5}{79626240}(191\Theta_{0,3}+60\Theta_{1,2}12573\theta_2^8-3384\theta_3^8+5553\theta_2^4\theta_3^4)\\	
	&+\frac{E_2^4}{38220595200}	
(9550\Theta_{0,4}-13100\Theta_{1,3}-14925\theta_2^8\theta_3^8+1257300 \theta_2^{12}\\	&-338400\theta_3^{12}+653850 \theta_2^8\theta_3^4-103500\theta_2^4\theta_3^8  +6783327\theta_2^8-2043648\theta_3^8+2553408 \theta_2^4\theta_3^4)\\
	&+\frac{E_2^3}{114661785600}( 1910\Theta_{0,5}-7550\Theta_{1,4}-1825\Theta_{2,3}+419100\theta_2^{16}-112800\theta_3^{16}\\
&-14250\theta_2^{12}\theta_3^4-239700\theta_2^4\theta_3^{12}-139050\theta_2^8\theta_3^8+6783327\theta_2^{12}-2043648\theta_3^{12}\\
&+3698703 \theta_2^8\theta_3^4-171072 \theta_2^4\theta_3^8)\\
&+\frac{E_2^2}{134842259865600}(+93590\Theta_{0,6}-695310 \Theta_{1,5}+841575\Theta_{2,4}+581875\theta_2^{12}\theta_3^{12}\\
&+30803850\theta_2^{20}-8290800\theta_3^{20}-37595250\theta_2^{16}\theta_3^4-31641750\theta_2^4\theta_3^{16}-11499075\theta_2^{12}\theta_3^8\\
&-6637050 \theta_2^8\theta_3^{12}
+997149069 \theta_2^{16}-300416256\theta_3^{16}+136431729\theta_2^{12}\theta_3^4\\
&-393132096 \theta_2^4\theta_3^{12}-154834659\theta_2^8\theta_3^8+3739608567\theta_2^{12}\\
& -602505216
\theta_3^{12}+1539809352\theta_2^8\theta_3^4+722608128\theta_2^4\theta_3^8)\\
&+\frac{E_2}{809053559193600}( 13370\Theta_{0,7}-157780\Theta_{1,6}+512295 \Theta_{2,5}\\
&-259700\Theta_{3,4}+6160770 \theta_2^{24}-1658160\theta_3^{24}-18724860\theta_2^{20} \theta_3^4\\
&-8921430\theta_2^4 \theta_3^{20}+12160575 \theta_2^{16} \theta_3^8+15038100\theta_2^8 \theta_3^{16}+5766075  \theta_2^{12} \theta_3^{12}\\
&+332383023\theta_2^{20}-100138752\theta_3^{20}-282160620\theta_2^{16}\theta_3^4-242867520\theta_2^4\theta_3^{16}-74943540 \theta_2^{12}\theta_3^8\\
&-53213265\theta_2^8\theta_3^{12}
+3739608567\theta_2^{16}-602505216\theta_3^{16}+382624803 \theta_2^{12}\theta_3^4\\
&-696066048\theta_2^4\theta_3^{12}-247053240 \theta_2^8\theta_3^8  ),\\
&+\frac{1}{155338283365171200}(26740\Theta_{0,8}-456400\Theta_{1,7}+2680300 \Theta_{2,6}\\
&-6175960\Theta_{3,5}+5432875\theta_2^{16}\theta_3^{16}\\
&+16428720\theta_2^{28}-4421760 \theta_3^{28}-90205080 \theta_2^{24} \theta_3^4-30140880 \theta_2^4 \theta_3^{24}+171302040 \theta_2^{20} \theta_3^8\\
&+155108520 \theta_2^8 \theta_3^{20}-72691500 \theta_2^{16} \theta_3^{12}-73353000 \theta_2^{12} \theta_3^{16}\\
&+1329532092 \theta_2^{24}-400555008\theta_3^{24}-3206708568\theta_2^{20} \theta_3^4\\
&-1375411968\theta_2^4 \theta_3^{20}+1608045642 \theta_2^{16} \theta_3^8+1957516092\theta_2^8 \theta_3^{16}+458968104 \theta_2^{12} \theta_3^{12}\\
&+29916868536\theta_2^{20}-4820041728\theta_3^{20}-20952823068\theta_2^{16}\theta_3^4-16917921792\theta_2^4\theta_3^{16}\\
&-5785406856\theta_2^{12}\theta_3^8-3716922816\theta_2^8\theta_3^{12}
+92227156875\theta_2^{16}).
\ea\ee

\addcontentsline{toc}{section}{References}

\bibliographystyle{utphys} 
\bibliography{Reference}

\end{document}